\documentclass[a4paper,preprintnumbers,floatfix,twocolumn,aps,prb,unsortedaddress,superscriptaddress]{revtex4-2}

\usepackage[para]{threeparttable} 
\usepackage{amsmath}
\usepackage{graphicx} 
\usepackage{booktabs}
\usepackage{subcaption}
\usepackage{url}
\usepackage{hyperref}
\usepackage{miller}
\usepackage{bm}
\usepackage{enumitem}
\usepackage{color}

\usepackage[paperwidth=210mm,paperheight=297mm,centering,hmargin=1.7cm,vmargin=2.5cm]{geometry}

\usepackage[utf8]{inputenc}

\begin{document}

\title{Modeling refractory high-entropy alloys with efficient machine-learned interatomic potentials: defects and segregation}

\author{J. Byggmästar}
\thanks{Corresponding author}
\email{jesper.byggmastar@helsinki.fi}
\affiliation{Department of Physics, P.O. Box 43, FI-00014 University of Helsinki, Finland}
\author{K. Nordlund}
\affiliation{Department of Physics, P.O. Box 43, FI-00014 University of Helsinki, Finland}
\author{F. Djurabekova}
\affiliation{Department of Physics, P.O. Box 43, FI-00014 University of Helsinki, Finland}
\affiliation{Helsinki Institute of Physics, Helsinki, Finland}

\date{\today}

\begin{abstract}
{We develop a fast and accurate machine-learned interatomic potential for the Mo--Nb--Ta--V--W quinary system and use it to study segregation and defects in the body-centred cubic refractory high-entropy alloy MoNbTaVW. In the bulk alloy, we observe clear ordering of mainly Mo--Ta and V--W binaries at low temperatures. In damaged crystals, our simulations reveal clear segregation of vanadium, the smallest atom in the alloy, to compressed interstitial-rich regions like radiation-induced dislocation loops. Vanadium also dominates the population of single self-interstitial atoms. In contrast, due to its larger size and low surface energy, niobium segregates to spacious regions like the inner surfaces of voids. When annealing samples with supersaturated concentrations of defects, we find that in complete contrast to W, interstitial atoms in MoNbTaVW cluster to create only small ($\sim 1$ nm) experimentally invisible dislocation loops enriched by vanadium. By comparison to W, we explain this by the reduced but three-dimensional migration of interstitials, the immobility of dislocation loops, and the increased mobility of vacancies in the high-entropy alloy, which together promote defect recombination over clustering.}
\end{abstract}

\maketitle

\section{Introduction}
\label{sec:intro}

High-entropy alloys are a new class of materials that is now being explored with increasing interest due to their many unique or enhanced properties, such as high mechanical and high-temperature strength, good resistance to corrosion, and enhanced radiation tolerance~\cite{cantor_microstructural_2004,tsai_high-entropy_2014,senkov_mechanical_2011,granberg_mechanism_2016,lu_enhancing_2016}. The majority of previous studies have focused on Fe- and Ni-based face-centred-cubic alloys~\cite{miracle_critical_2017}.  Considerably less is known about refractory alloys based on the body-centred-cubic group V and VI transition metals~\cite{senkov_development_2018}. With high melting points and mechanical strength, refractory metals and alloys are attractive for a variety of applications. For example, bcc metals are generally more tolerant to ion and neutron irradiation than fcc metals~\cite{zinkle_designing_2014}. This makes W-based high-entropy alloys particularly interesting for nuclear applications, such as the shielding material in fusion reactors~\cite{federici_european_2017}. Exceptional radiation tolerance was indeed recently demonstrated for a W-based high-entropy alloy, showing very little radiation hardening and no signs of radiation-induced dislocation loops even at high doses~\cite{el-atwani_outstanding_2019}.

The vast amount of possible alloy compositions makes the search for promising high-entropy alloys for a given application difficult. Computational modeling is therefore essential, both for guiding experimental manufacturing towards alloy compositions with the desirable properties and for understanding the atom-level mechanisms that give high-entropy alloys their unique properties. However, the chemical complexity also makes most modeling techniques challenging. Density functional theory (DFT) is an invaluable tool for quantifying fundamental material and defect properties, but is computationally too demanding for large-scale atomistic simulations. Beyond DFT, most modeling techniques rely on a parametrised model for the interatomic interactions, which for a high-entropy alloy with many elements involves exceedingly many cross-species interactions. Developing interaction models, like interatomic potentials or on-lattice cluster expansion models, for increasingly complex alloys is now a key step towards a more fundamental understanding of high-entropy alloys and their exotic properties. 

For W-based high-entropy alloys, recent studies along these directions have led to important insights. Cluster-expansion models and on-lattice machine-learning potentials have revealed strong preferential ordering and segregation at low temperatures in bulk W-based alloys~\cite{fernandez-caballero_short-range_2017-1,kormann_long-ranged_2017,kostiuchenko_impact_2019,liu_monte_2021}. Beyond on-lattice models, Li et al. recently developed a machine-learning potential for Mo--Nb--Ta--W alloys~\cite{li_complex_2020}. They used it to study the fundamental properties of screw and edge dislocations and found that Nb segregates to grain boundaries. However, not much effort has been focused on the quinary MoNbTaVW alloy. Here, we develop an accurate machine-learned interatomic potential for all Mo--Nb--Ta--V--W alloys. We use the potential to study the preferential ordering and segregation in bulk MoNbTaVW and around defects. Furthermore, we show how the mobility of vacancy and interstitial defects controls the defect evolution and leads to a vastly different defect structure compared to pure W.

\section{Methods}

\subsection{Machine-learning potential}
\label{sec:gap}

We first use the Gaussian approximation potential (GAP) framework~\cite{bartok_gaussian_2010,bartok_gaussian_2015} to train an interatomic potential for the complete Mo--Nb--Ta--V--W system. With five elements and the aim to get a fairly general potential that is at least reasonably transferable to any alloy composition, constructing the training database is a challenge. We also require that the potential can describe not only bulk crystals, but also any form of defects that may be present or form due to irradiation. We found that with a reasonably sized training data set, using the many-body SOAP descriptor~\cite{bartok_representing_2013} that is typically used in GAPs leads to overfitting issues (showing significantly larger test errors than training errors). Converging a SOAP-GAP towards DFT accuracy for independent test data would likely require a prohibitively large training set due to the vast descriptor space with five elements. Our solution is therefore to rely on low-dimensional two- and three-body descriptors, which require less data but as we demonstrate can still achieve remarkably good accuracy for multicomponent alloys. In fact, it outperforms a SOAP-GAP trained to the same training data. Additionally, the use of only low-dimensional descriptors allows for a tabulation scheme that provides a significant computational speedup, as discussed in Sec.~\ref{sec:tabgap}.

We use the following expression for the total energy of a system of $N$ atoms
\begin{align}
 \begin{split}
    E_\mathrm{tot.}^\mathrm{GAP} &= \sum_{i<j}^N E_{ij}^\mathrm{rep.} (r_{ij}) + \delta_\mathrm{2b}^2 \sum_{i<j}^N \sum_p^{M_\mathrm{2b}} \alpha_{ij, p} K_\mathrm{2b} (r_{ij}, r_p) \\
    &+ \delta_\mathrm{3b}^2 \sum_{i, j<k}^N \sum_t^{M_\mathrm{3b}} \alpha_{ijk, t} K_\mathrm{3b} (\bm{q}_{ijk}, \bm{q}_t).
  \end{split}
  \label{eq:gap}
\end{align}
Here, the first term is a purely repulsive screened Coulomb potential fitted to each element pair using all-electron DFT data~\cite{nordlund_repulsive_1997}. Our methods for fitting and including the repulsive pair potentials are described in detail in Ref.~\cite{byggmastar_machine-learning_2019}. The second term is a machine-learning pair potential, carried out as Gaussian process regression with the interatomic distance of each pair $ij$ as the descriptor. The final term is the three-body machine-learning term as a sum over all atom triplets $ijk$. $\alpha$ are the optimised regression coefficients, and $K_\mathrm{2b}$, $K_\mathrm{3b}$ are the kernel functions, for which we use the squared-exponential kernel with 1 Å as the length-scale hyperparameter. $M_\mathrm{2b}$ and $M_\mathrm{3b}$ are the number of representative pairs $p$ and triplets $t$ from the training structures used to carry out the regression. We use $M_\mathrm{2b}=20$ and $M_\mathrm{3b}=300$ for all element combinations. Values higher than $M_\mathrm{3b}=300$ provided similar accuracy, but showed an increasing (although small) tendency of overfitting. The machine-learning energy predictions are scaled by the parameters $\delta_\mathrm{2b}^2 = 10$ eV and $\delta_\mathrm{3b}^2 = 1$ eV. The GAP framework and its parameters are described in more detail in e.g. Refs.~\cite{bartok_gaussian_2015,bartok_machine_2018}.

The three-body descriptor calculated for each triplet of atoms $ijk$ is the vector~\cite{bartok_gaussian_2015}
\begin{equation}
    \bm{q}_{ijk} =
    \begin{pmatrix}
    r_{ij} + r_{ik} \\
    (r_{ij} - r_{ik})^2 \\
    r_{jk} \\
    \end{pmatrix}
    f_\mathrm{cut}(r_{ij}) f_\mathrm{cut}(r_{ik}),
\end{equation}
which is invariant to permutations of $jk$. Smooth cutoff functions $f_\mathrm{cut}(r)$ are imposed on the bonds to the central atom $i$. We use 5 Å cutoff radii for both the two-body and three-body descriptors and all element pairs and triplets. Initially, we experimented with longer cutoffs for the pairwise potentials (to capture possible long-range interactions) and shorter three-body cutoffs, but found that 5 Å cutoffs for both descriptors lead to better accuracy.

The potential is trained to total energies, forces, and virial stresses obtained from DFT calculations using \textsc{vasp}~\cite{kresse_ab_1993,kresse_ab_1994,kresse_efficiency_1996,kresse_efficient_1996} for a variety of Mo--Nb--Ta--V--W structures. The training set includes bulk Mo--Nb--Ta--V--W bcc crystals sampled at all compositions, including the pure elements. The majority of the alloys are randomly ordered, but we also found it necessary to include ordered alloys. Defected structures (up to 5 vacancies and self-interstitial atoms) are only included in the pure elements and in equiatomic random MoNbTaVW alloys. Liquid structures are included for pure elements and all equiatomic binary, ternary, quaternary, and quinary alloys of Mo--Nb--Ta--V--W, but with most structures for the equiatomic MoNbTaVW high-entropy alloy (HEA). The training set additionally includes HEA crystals with one randomly added interstitial atom that is relatively close (but not too close) to its neighbour atom. These structures ensure that the repulsion between all elements in crystals is captured. Surface structures are included for the pure elements along with a few HEA surfaces to ensure some transferability to alloy surfaces, although we do not specifically target surface properties. Our strategy for constructing the training structures and the training process is described in more detail in the Supplemental material online. The potential is trained using the \textsc{quip} code (\url{https://github.com/libAtoms/QUIP}).

\subsection{Tabulated machine-learning potential}
\label{sec:tabgap}

Carrying out the Gaussian process regression terms of Eq.~\ref{eq:gap} is computationally expensive due to the loop over the (sparsified) training points $M_\mathrm{2b}$, $M_\mathrm{3b}$. Because our GAP only contains pure two- and three-body-dependent terms, it is possible to create a tabulated version of the potential, which we will call tabGAP. Creating computationally efficient tabulated machine-learning potentials was initially demonstrated by Glielmo et al.~\cite{glielmo_efficient_2018} and recently further developed in the \textsc{flare} framework~\cite{vandermause_--fly_2020,xie_bayesian_2021}, although details vary from our approach. The pairwise terms can be evaluated as 1D interpolations between tabulated pair energies and the three-body terms as 3D interpolations between tabulated triplet energies. We use a grid of $(r_{ij}, r_{ik}, \cos \theta_{ijk})$ points for the three-body interpolation. With $S^\mathrm{1D}$ and $S^\mathrm{3D}$ representing 1D and 3D cubic spline interpolations, the total energy of the tabGAP becomes
\begin{equation}
    E_\mathrm{tot.}^\mathrm{tabGAP} = \sum_{i<j}^N S_{ij}^\mathrm{1D}(r_{ij}) + \sum_{i, j<k}^N S_{ijk}^\mathrm{3D} (r_{ij}, r_{ik}, \cos \theta_{ijk}),
    \label{eq:tabgap}
\end{equation}
where the repulsive pair potential and the GAP pair potential in Eq.~\ref{eq:gap} are merged into one 1D spline per element pair. With five elements, there are 15 unique pairs and 75 triplets (with symmetry on the $jk$ elements), each requiring a tabulated grid of energies. With sufficiently dense grids, the tabGAP is virtually identical to the original GAP. The convergence of the interpolation error with increasing numbers of grid points is demonstrated in the Supplemental material online.

We have implemented this cubic-spline-interpolated potential as a \texttt{pair\_style} for \textsc{lammps}~\cite{plimpton_fast_1995} (based on the implementation from the open-source \textsc{flare} code~\cite{vandermause_--fly_2020,xie_bayesian_2021}). The speed-up compared to the original GAP evaluated as in Eq.~\ref{eq:gap} is around two orders of magnitude with the current implementation, making it comparable in speed to traditional angular-dependent potentials like Tersoff and MEAM potentials~\cite{tersoff_new_1988,baskes_application_1987,lee_second_2000}. The potential file is available from~\cite{byggmastar_ida_tabgap_2021} and the tabulation and \textsc{lammps} implementation from \url{https://gitlab.com/jezper/tabgap}.

\subsection{Simulation methods}

All DFT calculations are carried out with \textsc{vasp}~\cite{kresse_ab_1993,kresse_ab_1994,kresse_efficiency_1996,kresse_efficient_1996} with PAW potentials~\cite{blochl_projector_1994,kresse_ultrasoft_1999} (\texttt{\_pv} for Ta and \texttt{\_sv} for all other elements), the PBE GGA exchange-correlation functional~\cite{perdew_generalized_1996}, a 500 eV cutoff energy, 0.15 Å$^{-1}$ maximum $k$-point spacing on $\Gamma$-centred Monkhorst-Pack grids~\cite{monkhorst_special_1976}, and 0.1 eV Methfessel-Paxton smearing~\cite{methfessel_high-precision_1989}. All molecular statics and dynamics simulations are done with \textsc{lammps}~\cite{plimpton_fast_1995} with a custom implementation of the tabulated machine-learned potential (tabGAP) as discussed above.

The average lattice constant and mixing energy of bulk equiatomic HEA properties are obtained by relaxing 50 different randomly ordered 2000-atom bcc systems. The simulation cells were cubic during the energy and pressure minimization. Allowing for noncubic relaxation produced cubic systems within the statistical uncertainty of the cubic bcc lattice constant. The elastic constants are computed for the same relaxed 50 HEA systems. The bulk modulus is obtained from volume-energy fits to the Birch-Murnaghan equation of state~\cite{birch_finite_1947}. The remaining elastic constants are solved from parabolic fits to strain-energy curves of the tetragonal and trigonal deformation modes (see e.g. Ref.~\cite{holec_trends_2012}), using $\pm 2\%$ strain intervals and optimising the atom positions at every strain.

Nudged elastic band calculations for vacancy migration barriers are performed in boxes of 128 atoms using \textsc{lammps}. The migration energy is obtained from the saddle point of the converged barrier. For relaxing single self-interstitials we use boxes of 251 atoms. For the formation energies of single vacancies we use 54-atom boxes in both \textsc{vasp} and \textsc{lammps}. We optimise both the positions and relax the box size to zero pressure for the vacancy and self-interstitial calculations. The formation energy for an $A$ vacancy (where $A$ is Mo, Nb, Ta, V, or W) is calculated as
\begin{equation}
    E_\mathrm{f}^{A\mathrm{vac}} = E_\mathrm{vac} - E_\mathrm{bulk} + \mu_A,
\end{equation}
where $E_\mathrm{vac}$ is the total energy of the relaxed vacancy system and $E_\mathrm{bulk}$ is the total energy of the relaxed HEA bulk with the vacancy filled by an atom of element $A$. The chemical potential $\mu_A$ is approximated simply as the energy per atom of $A$ in its ground state (bcc bulk), which is close to the real chemical potential estimated using other methods~\cite{zhao_defect_2020}. The vacancy relaxation volume is calculated as
\begin{equation}
    \Omega_\mathrm{rel.}^{A\mathrm{vac}} = \frac{V_\mathrm{vac} - V_\mathrm{bulk}}{\Omega_A},
\end{equation}
where $V_\mathrm{vac}$ and $V_\mathrm{bulk}$ are the total volumes of the relaxed vacancy and bulk HEA systems and $\Omega_A$ is the volume per atom of pure bcc $A$.

Relaxing self-interstitial atoms in the HEA often results in stable interstitial configurations far from the initial configuration, as many local atomic environments provide no local minima. This, combined with the fact that many interstitial configurations are mixed dumbbells, makes the choice or construction of the corresponding reference bulk system somewhat ambiguous. Hence, we define and compute the formation energy of an $A$--$B$ interstitial dumbbell configuration as the formation energy of the entire interstitial system with $N = 251$ atoms ($E_\mathrm{f}^\mathrm{SIA}$), and then compare it to an average defect-free HEA bulk reference to get only the formation energy associated with the interstitial as
\begin{equation}
    E_\mathrm{f}^{A\text{-}B} = E_\mathrm{f}^\mathrm{SIA} - N E_\mathrm{mix}^\mathrm{HEA}.
\end{equation}
The formation energy and mixing energy (per atom) of a HEA system of $N$ atoms with total energy $E$ is $E_\mathrm{f} = E - \sum_i^N E_i$ and $E_\mathrm{mix} = E_\mathrm{f} / N$, where $E_i$ is the energy per atom of the pure ground state of atom $i$. $E_\mathrm{mix}^\mathrm{HEA}$ is computed as the average mixing energy of randomly ordered HEA systems and is $-41.85$ meV/atom in the tabGAP (as listed in Tab.~\ref{tab:bulk} and discussed later).

Hybrid Monte Carlo--Molecular dynamics (MC+MD) simulations are carried out as implemented in \textsc{lammps}. MD is done in the $NPT$ ensemble at different constant temperatures and zero pressure. For the single-crystal simulations we use 100 trial MC atom swaps every 10 MD steps using boxes of 6750 atoms. The simulations are continued until the potential energy and short-range order parameters have sufficiently converged, although for at least 1 million MC steps and about 8 million MC steps for the lower temperatures. The short-range order parameter for element pairs $AB$ with interatomic bonds in the interval $\Delta r_{ij}$ is computed using the definition
\begin{equation}
 S_{\Delta r_{ij}}^{AB} = 1 - \frac{p_{\Delta r_{ij}}^{AB}}{c_B},
\end{equation}
where $p_{\Delta r_{ij}}^{AB}$ is the probability of finding a $B$ atom around $A$ in the neighbour shell $\Delta r_{ij}$ and $c_B$ is the concentration of $B$ in the alloy. For first-nearest-neighbour pairs, we used the interval $\Delta r_{ij} = [0, 3]$ Å and for the second-nearest neighbours $\Delta r_{ij} = [3, 3.9]$ Å.

For the void and dislocation loop MC+MD simulations, we allow for more MD relaxation with 10 MC trial swaps every 10 MD steps in boxes of around 20 000 atoms. Here, we run the simulations until the concentrations around the defects have stabilised (disregarding the surrounding bulk), which only required a few hundred thousand MC steps.

The defect annealing simulations are done using boxes of 250 000 atoms with 10 000 randomly created Frenkel pairs, corresponding to a supersaturated (4\%) defect concentration. The atomic positions are first optimised, which already annihilates many Frenkel pairs and causes some initial clustering of interstitials and vacancies. This is followed by a 1 ns $NPT$ annealing run at 2000 K and zero pressure. The final frames are then again optimised to allow a more reliable analysis of the final defect structure. The dislocations are identified using the dislocation extraction algorithm (DXA)~\cite{stukowski_automated_2012} in \textsc{ovito}~\cite{stukowski_visualization_2010}. Interstitials and vacancies are found using the Wigner-Seitz method and grouped into clusters with the cutoff radius between the second and third nearest neighbour for vacancies and between the third and fourth for interstitials.

\section{Results}

\subsection{Validating the machine-learned potential}
\label{sec:validation}

\begin{table}
    \caption{Root-mean-square energy ($E$) and force component ($F$) errors of the tabGAP compared to DFT for independent test sets of different classes of structures. $N_\mathrm{s}$ is the number of structures in the test sets (although the number of atoms in different structure classes vary).}
    \label{tab:rmse}
    \centering
    \begin{tabular}{lccc}
        \toprule
        Structure type & $N_\mathrm{s}$ & $E$ (meV/atom) & $F$ (eV/Å) \\
        \midrule
        Bulk pure elements & 220 & 3.17 & 0.15 \\
        Bulk random alloys & 100 & 2.99 & 0.11 \\
        Bulk ordered alloys & 165 & 3.31 & 0.04 \\
        Bulk HEA & 40 & 3.03 & 0.09 \\
        Vacancies in HEA & 10 & 3.59 & 0.09 \\
        Interstitials in HEA & 10 & 3.39 & 0.22 \\
        Liquid pure elements & 225 & 28.61 & 0.71 \\
        Liquid HEA & 11 & 40.03 & 0.78 \\
        \bottomrule
    \end{tabular}
\end{table}

\begin{figure*}
    \centering
    \includegraphics[width=0.85\linewidth]{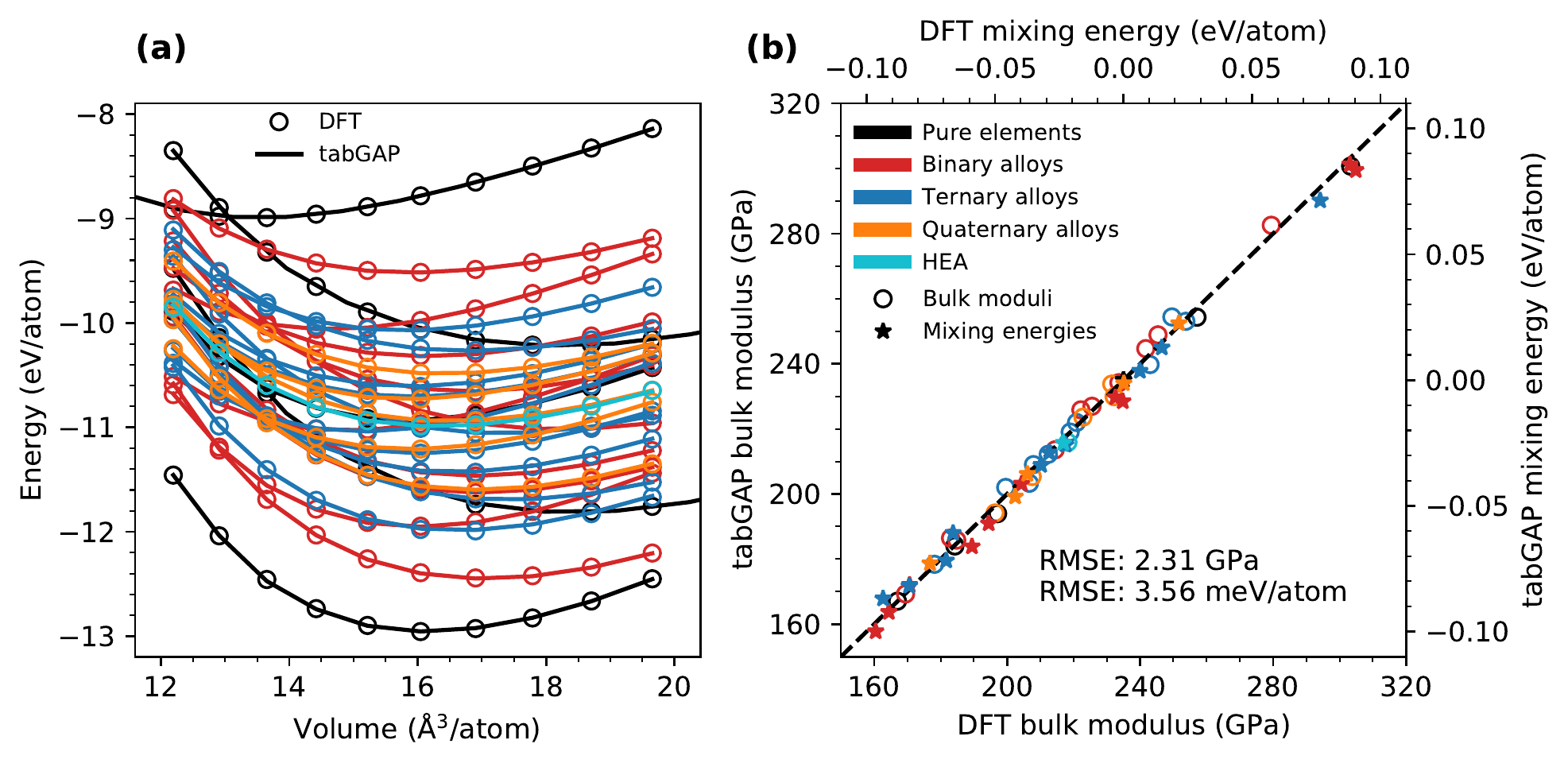}
    \includegraphics[width=0.85\linewidth]{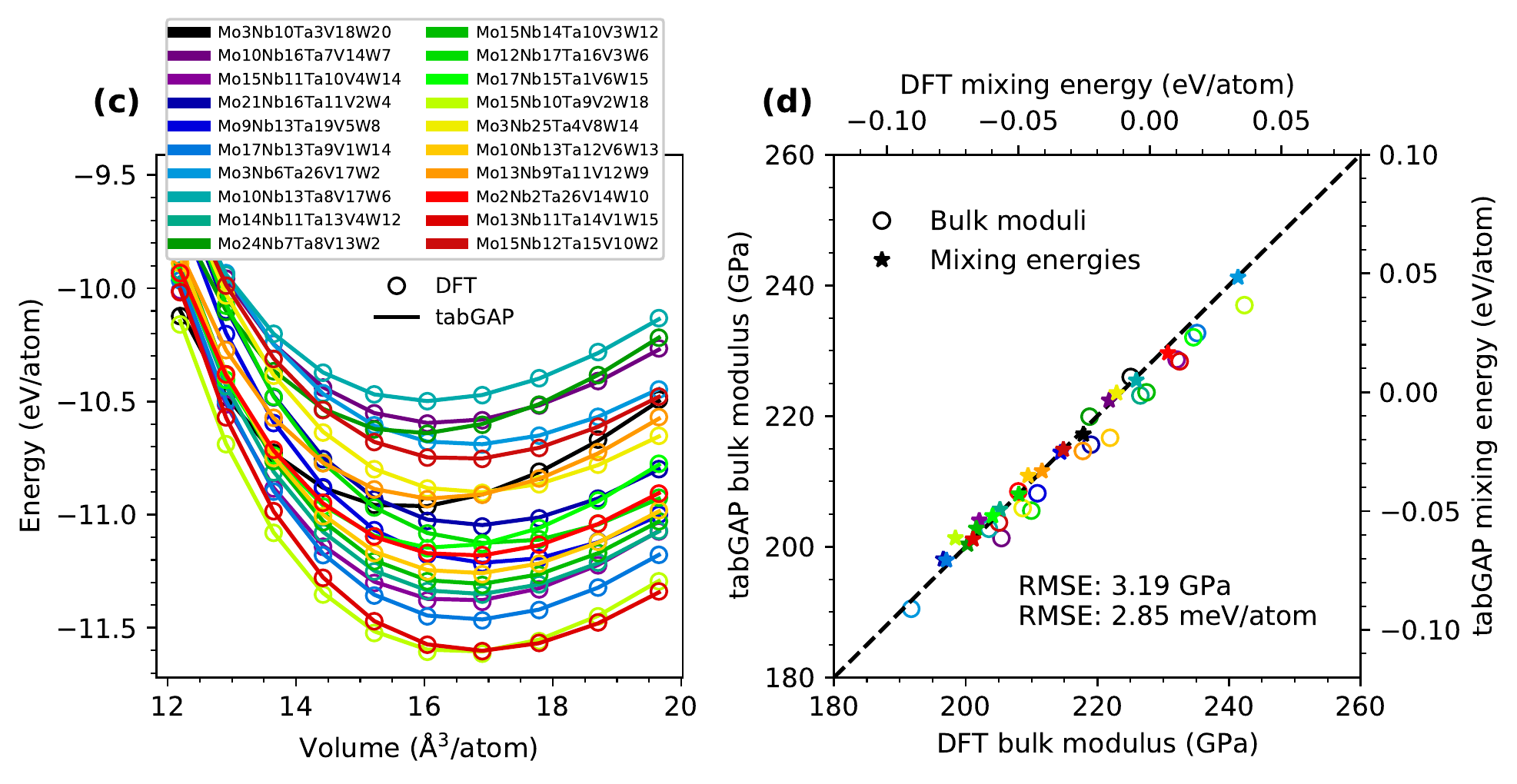}
    \caption{Comparison between DFT and the tabGAP for bulk moduli and mixing energies for a test set of Mo--Nb--Ta--V--W alloys. Energy per atom as a function of volume for random alloys for (a) all equiatomic compositions and (c) 20 different randomly sampled Mo-Nb-Ta-V-W alloy compositions. (b), (d): Bulk moduli and mixing energies at equilibrium volumes obtained from the energy--volume data and compared between DFT and the tabGAP.}
    \label{fig:ev-Bemix}
\end{figure*}

Before using the tabGAP to study segregation and the defect structure of the equiatomic MoNbTaVW high-entropy alloy, we here briefly demonstrate the accuracy of the potential. Table~\ref{tab:rmse} lists root-mean-square errors (RMSE) of the tabGAP compared to DFT for a range of structures used as test sets. All atoms in the bulk crystals have been randomly displaced from the perfect lattice positions to produce significant interatomic forces. The accuracy of the tabGAP is around 3 meV/atom and 0.05--0.2 eV/Å for any given composition of Mo--Nb--Ta--V--W as well as for HEA lattices containing vacancies or self-interstitial atoms. Some further test results are provided in the Supplemental material, where we show that the tabGAP also reproduces the W--Mo alloy training dataset from Ref.~\cite{nikoulis_machine-learning_2021} with similar accuracy as in Tab.~\ref{tab:rmse}, verifying that the potential is accurate also for the binary alloys.

To further verify that the tabGAP can describe the phase stability and elastic response of arbitrary alloy compositions, we compare the 0 K equation of state for a variety of alloys between DFT and the tabGAP in Fig.~\ref{fig:ev-Bemix}. From the energy-volume data, we obtain the bulk moduli and equilibrium mixing energies from fits to the Birch-Murnaghan equation of state~\cite{birch_finite_1947}. Fig~\ref{fig:ev-Bemix}(a) shows the energy-volume data for randomly ordered alloys at all equiatomic alloy compositions (one system per composition) and Fig.~\ref{fig:ev-Bemix}(c) shows data for 20 different quinary alloys randomly ordered and at randomly sampled compositions. The corresponding bulk moduli and mixing energies are shown in Figs.~\ref{fig:ev-Bemix}(b) and (d), compared between DFT and the tabGAP. The tabGAP reproduces the DFT data within only a few meV/atom for mixing energies and a few GPa for the bulk moduli. Also noteworthy is that the agreement between DFT and the tabGAP is good for the entire wide range of volumes for all alloys in Figs.~\ref{fig:ev-Bemix}(a) and (c), where each curve spans an energy difference around 1--2 eV/atom.

In table~\ref{tab:bulk}, we list average properties of MoNbTaVW predicted by the tabGAP and computed from 50 relaxed 2000-atom systems. The lattice constant in the tabGAP (3.195 Å) is close to the experimental value 3.1832 Å~\cite{senkov_refractory_2010}.

\begin{table}
    \caption{Bulk properties of MoNbTaVW at 0 K obtained with the tabGAP. The values are the averages and standard deviations of 50 randomly ordered and relaxed 2000-atom systems.}
    \label{tab:bulk}
    \centering
    \begin{tabular}{lr}
        \toprule
         & MoNbTaVW \\
        \midrule
        $a$ (Å) & $3.195 \pm 0.001$ Å \\
        $E_\mathrm{mix}$ (meV/atom) & $-41.85 \pm 0.71$ \\
        $B$ (GPa) & $210.4 \pm 0.3$ \\
        $C_{11}$ (GPa) & $382.2 \pm 0.6$ \\
        $C_{12}$ (GPa) & $124.5 \pm 0.3$ \\
        $C_{44}$ (GPa) & $47.5 \pm 0.3$ \\
        \bottomrule
    \end{tabular}
\end{table}

Reproducing liquid properties for arbitrary compositions, densities, and temperatures with very high accuracy cannot be expected with a simple three-body-dependent potential. As seen in Tab.~\ref{tab:rmse}, the RMSEs for liquids are up to 10 times higher than for any crystalline structures. Nevertheless, we confirmed that the tabGAP still provides a reasonable descripton of the liquid phase and melting. The melting point of the HEA predicted by the tabGAP is $2760 \pm 20$ K as determined by $NPT$ molecular dynamics (MD) simulations of a solid-liquid interface. A rough estimate of the experimental melting point can be taken as the average of the pure-element melting temperatures, yielding 2961 K~\cite{rumble_crc_2019}. However, previously we found that even highly accurate many-body GAPs tend to underestimate the melting point slightly~\cite{byggmastar_gaussian_2020}. The corresponding average of the pure-element many-body GAPs from Ref.~\cite{byggmastar_gaussian_2020} is 2796 K~\cite{byggmastar_gaussian_2020}, which is very close to the HEA melting point predicted by the tabGAP.

\subsection{Order and disorder in bulk MoNbTaVW}
\label{sec:sro}

\begin{figure*}
    \centering
    \includegraphics[width=0.35\linewidth]{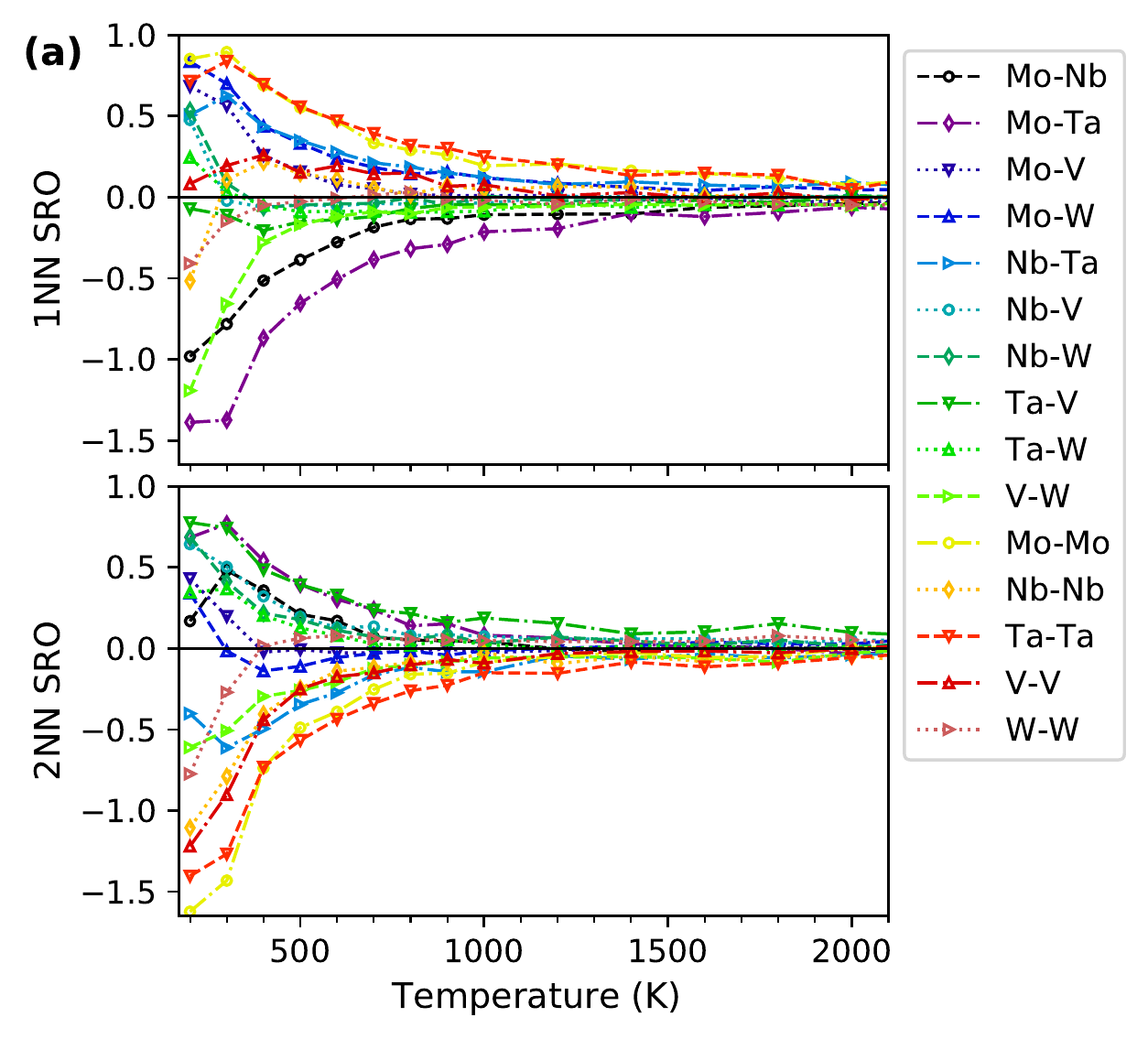}
    \includegraphics[width=0.21\linewidth]{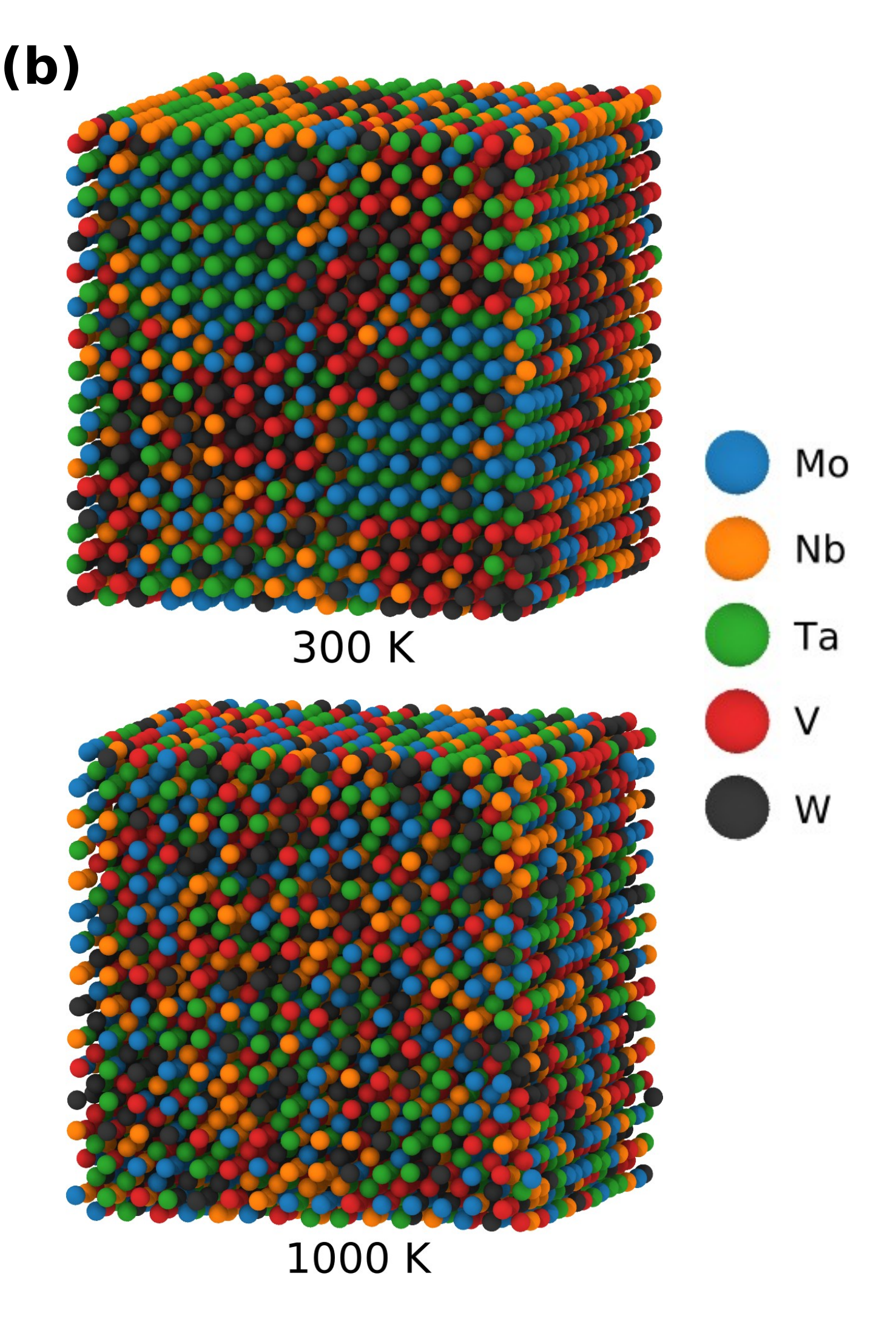}
    \includegraphics[width=0.35\linewidth]{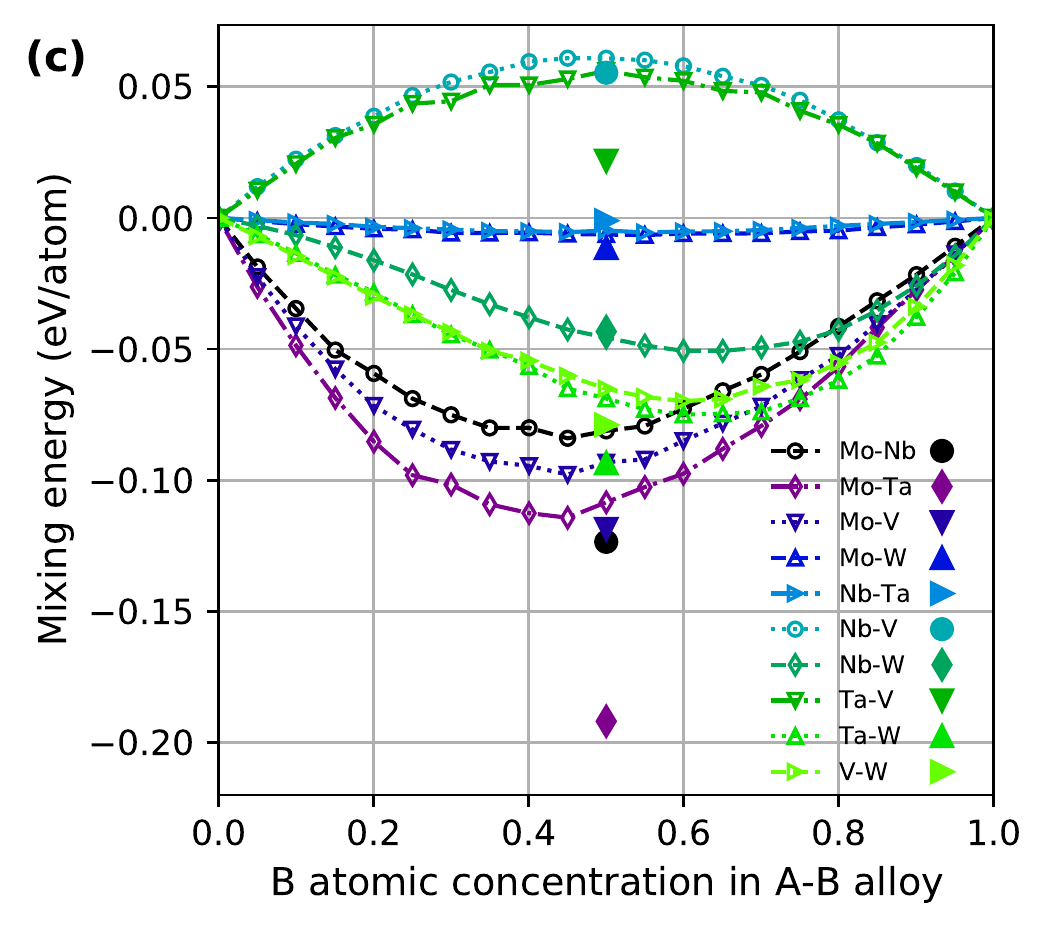}
    \caption{Short-range order in MoNbTaVW. (a): Short-range order parameters as a function of temperature from the MC+MD simulations for both first-nearest-neighbour (1NN) and 2NN pairs. (b): Snapshots of the final lattices at two temperatures, showing clear local ordering at 300 K. (c): Mixing energy curves for randomly ordered binary alloys. The large solid points are the ordered CsCl (B2) phase.}
    \label{fig:bulk-segreg}
\end{figure*}

We first apply the tabGAP in a study of ordering and segregation in the single-crystal equiatomic HEA using hybrid MC+MD simulations. Previous studies using cluster expansion models and DFT calculations have established that there is a strong preference for local ordering of Mo--Ta binary crystals due to their relatively low mixing energy~\cite{widom_hybrid_2014,fernandez-caballero_short-range_2017-1,yin_ab_2020}. Additionally, it has been previously shown that including lattice relaxation is crucial in order to reproduce the correct phase stabilities at finite temperatures~\cite{kostiuchenko_impact_2019}. Relaxation effects and realistic thermal vibrations are here naturally included in the $NPT$ MC+MD simulations with the tabGAP. Since the tabGAP is trained to both random and ordered alloys, including all the binary alloys and MC+MD-optimised ordered systems obtained in an active-learning fashion (see the Supplemental material), we are confident that it can accurately model order and disorder in the HEA.

Fig.~\ref{fig:bulk-segreg} shows the results from the MC+MD simulations. The 1NN and 2NN short-range order parameters as functions of temperature in Fig.~\ref{fig:bulk-segreg}(a) show, consistent with previous studies, that Mo and Ta are locally ordered at low temperatures. A strongly negative 1NN Mo--Ta short-range order (SRO) value and a positive 2NN SRO value indicate the presence of MoTa in the CsCl (B2) order, which is confirmed by visually analysing the lattice. Fig.~\ref{fig:bulk-segreg}(b) shows snapshots of the systems at two temperatures. In addition to Mo--Ta ordering, Mo--Nb and W--V 1NN pairs are also favoured. W--V pairs stand out by also having negative 2NN SRO value, indicating a more complex structure than the CsCl order, as was also observed in Ref.~\cite{fernandez-caballero_short-range_2017-1}. All SRO values start approaching zero after 300 K, initiating the transition from the ordered phases to a random solid solution. Still, a relatively strong local ordering of mainly Mo--Ta is retained up to temperatures beyond 1000 K.

The preferential ordering and segregation can largely be understood by the mixing energies of the binary alloys. Fig.~\ref{fig:bulk-segreg}(c) shows mixing curves at 0 K for all binaries as given by the tabGAP (which are identical within 2.9 meV/atom to DFT data, see figure S2 in the Supplemental material). The connected data points in Fig.~\ref{fig:bulk-segreg}(c) are mixing energies for randomly ordered alloys (using 1024-atom boxes) and the large solid points are the ordered CsCl phase. The mixing curves show that Mo--Ta alloys have the lowest mixing energies and that the ordered MoTa phase is by far the most preferred binary alloy. The mixing energies also suggest that Nb--V alloys are the least favoured, with clearly positive mixing energies for both random and the ordered CsCl phase. This is reflected in the MC+MD simulations by the fact that Nb and V are strongly segregated from each other, showing positive SRO values for both 1NN and 2NN pairs.

\subsection{Vacancies and self-interstitial atoms in MoNbTaVW}
\label{sec:vac_int}

\begin{figure}
    \centering
    \includegraphics[width=\linewidth]{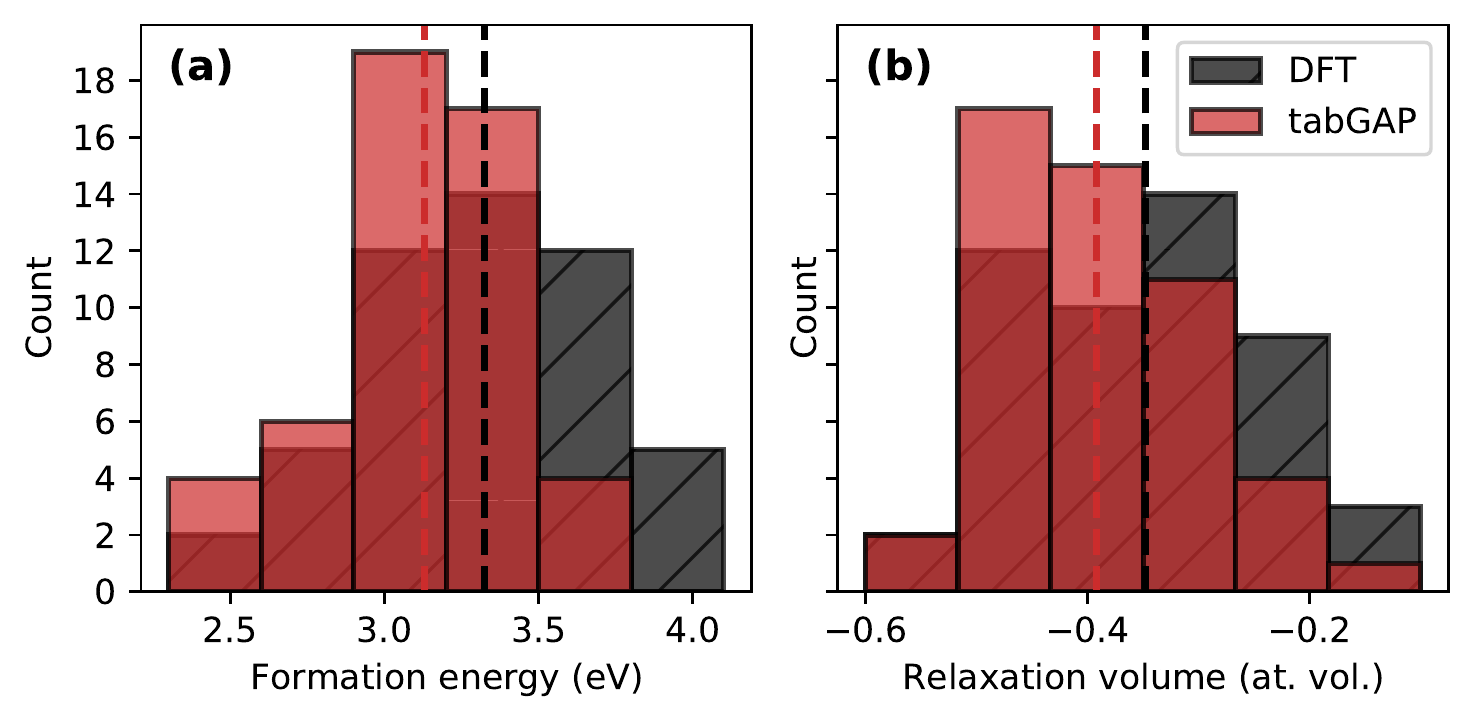}
    \includegraphics[width=\linewidth]{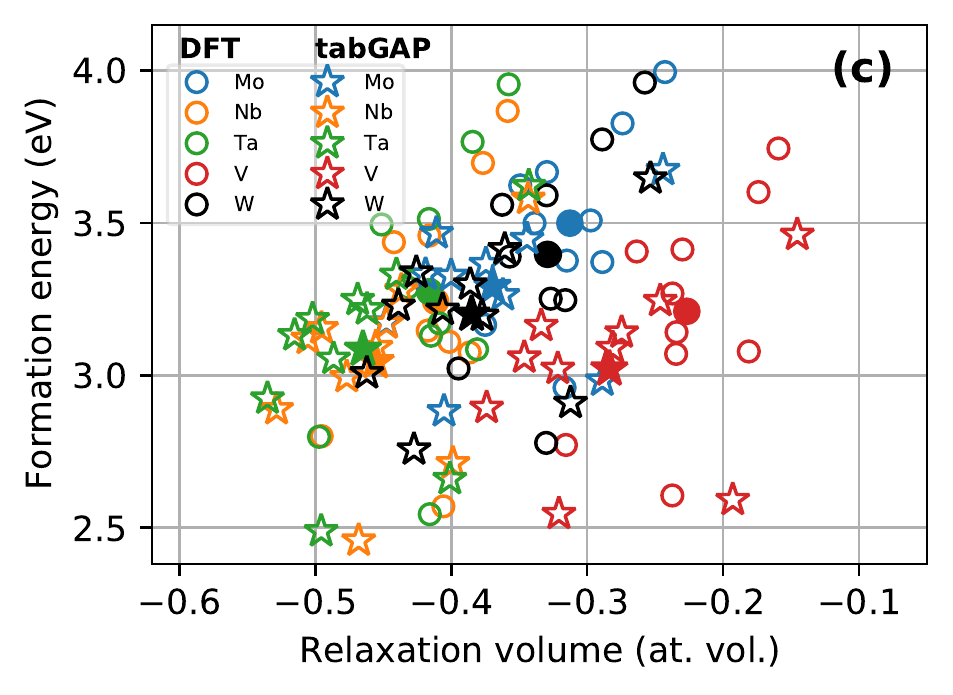}
    \caption{Formation energies and relaxation volumes of single vacancies MoNbTaVW. (a) Distribution of formation energies and (b) distribution of relaxation volumes compared between DFT and the tabGAP. Dashed lines are the averages. (c) The same data separated by element (of the filled vacancy in the reference bulk). Solid points are averages.}
    \label{fig:vac_ef}
\end{figure}

The presence of defects introduce local lattice relaxation and long-range strain fields that may influence the preferential ordering and segregation of the different-sized atoms in the HEA. We use the tabGAP to investigate the energetics of single vacancies and self-interstitials in the HEA. The chemical complexity provides exceedingly many possible configurations for even these simple defects, which calls for a statistical treatment. Here, we only consider randomly ordered HEAs with randomly added single vacancies and interstitials, followed by relaxation. Fig.~\ref{fig:vac_ef}(a)--(b) shows distributions of formation energies and relaxation volumes of single vacancies in the HEA. The vacancies are created in 10 different HEA systems. For each vacancy system, reference bulk systems are created by filling the vacancy with each element separately, so that we in total obtain data for 50 different vacancies. Fig.~\ref{fig:vac_ef} also shows results from DFT calculations to further validate the predictive accuracy of the tabGAP. The tabGAP slightly underestimates the average formation energy and the relaxation volume compared to DFT, but overall provides similar distributions. The average vacancy formation energy is 3.3 eV according to DFT and 3.1 eV according to the tabGAP. The average relaxation volume is $-0.35$ at. vol. in DFT and $-0.40$ at. vol. in the tabGAP.

The single-vacancy data are separated by element in Fig.~\ref{fig:vac_ef}(c), revealing additional insight. First, there is no correlation in the formation energy of vacancies between the elements, with energies in the range 2.5--4 eV for all elements. There is, however, a clear separation between the elements for relaxation volumes. Filling a vacancy with a V atom, the smallest atom, cause the least relaxation around the vacancy with most relaxation volumes around $-0.2$ to $-0.3$ at. vol. Vice versa, being the largest atoms, Nb and Ta vacancies produce significantly stronger relaxation with relaxation volumes around $-0.4$ to $-0.5$ at. vol.

\begin{figure}
    \centering
    \includegraphics[width=\linewidth]{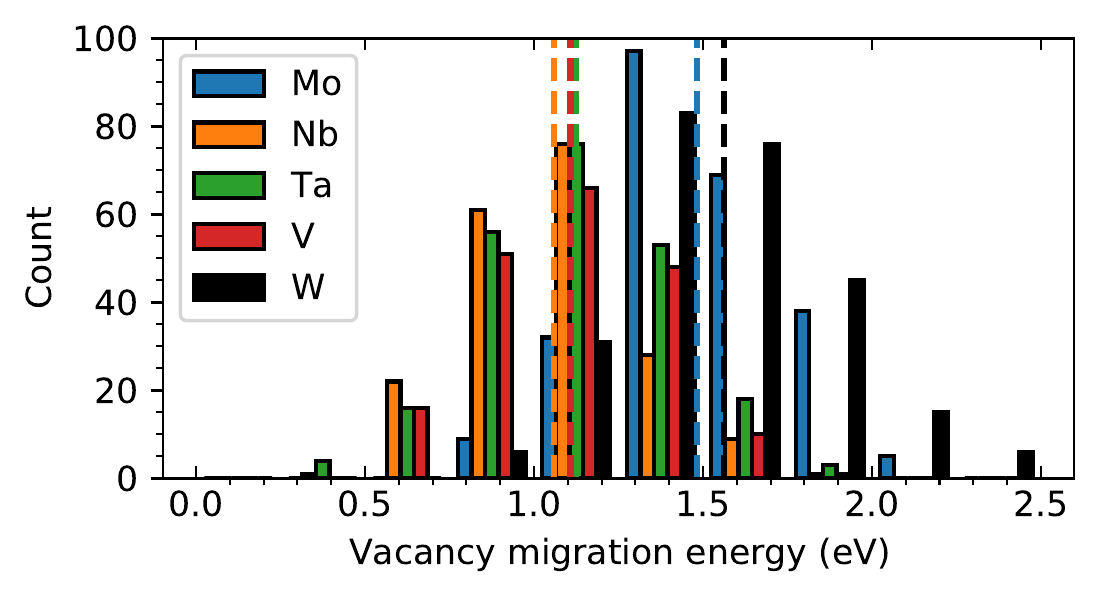}
    \caption{Distribution of vacancy migration energies in MoNbTaVW. The dashed lines are the averages for each migrating element.}
    \label{fig:vac_mig}
\end{figure}

Fig.~\ref{fig:vac_mig} shows distributions of migration energies for single vacancies in the HEA. In total, we calculated over 1100 migration barriers in different randomly ordered HEAs with the tabGAP to obtain reliable statistics. Fig.~\ref{fig:vac_mig} reveals that the migration energies for each element are roughly normally distributed and span a wide energy range of around 1 eV. There is again a clear separation between the elements of the migrating atom. The group 6 elements W and Mo have the highest migration energies. The average migration energy for W is 1.56 eV and for Mo 1.48 eV. The group 5 elements show significantly lower migration energies, with the averages 1.06 eV for Nb, 1.12 eV for Ta, and 1.11 eV for V. The trend is consistent with the vacancy migration energies in the pure elements, where W and Mo have by far the highest migration energies (around 1.7 and 1.3 eV~\cite{ma_effect_2019}) while the group 5 elements all have migration energies around 0.6--0.7 eV~\cite{ma_effect_2019}. The result of alloying on the vacancy migration is thus a wide distribution of migration energies where the averages are reduced for W and Mo and increased for Nb, Ta, and V.

\begin{figure}
    \centering
    \includegraphics[width=\linewidth]{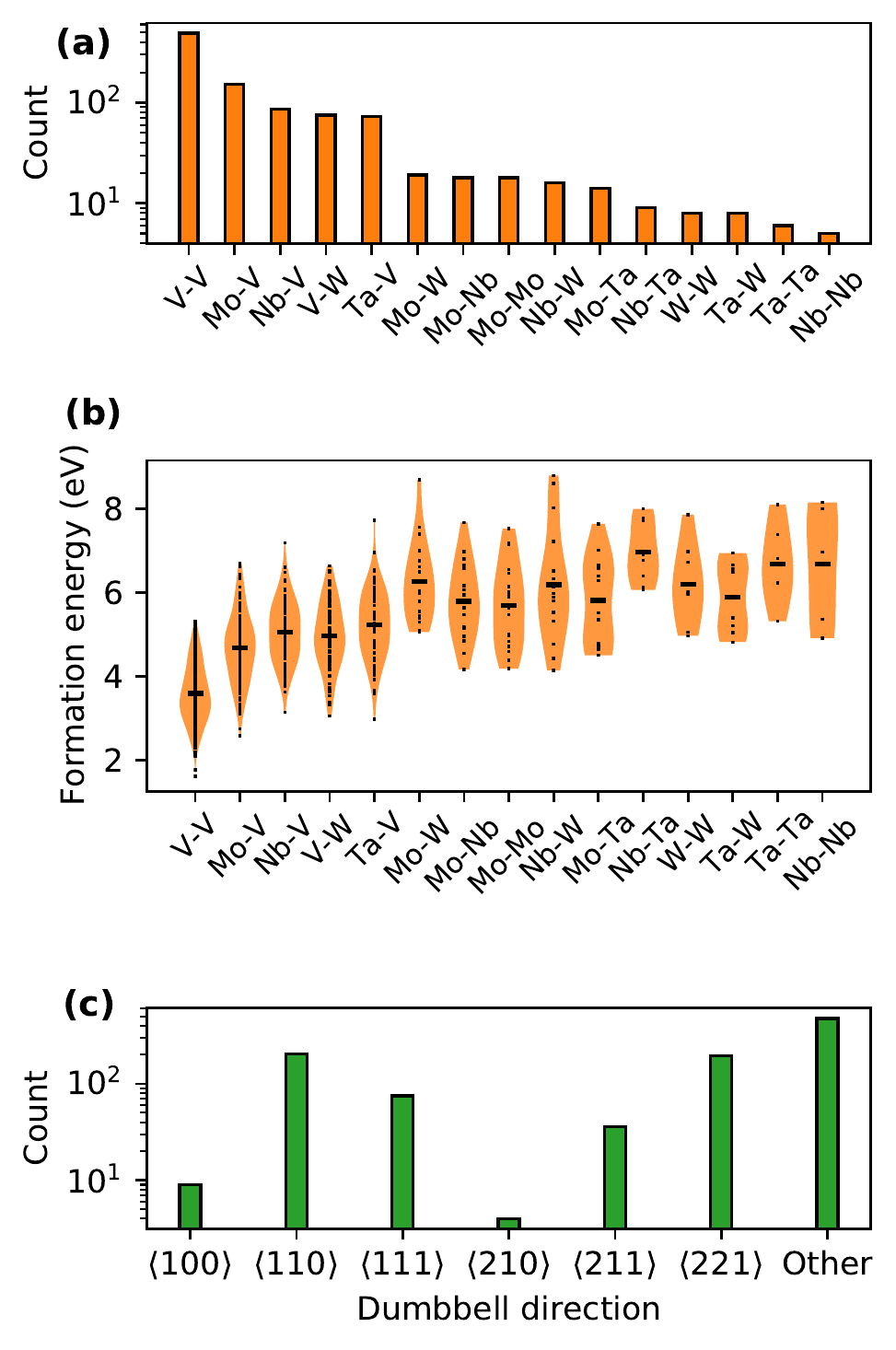}
    \caption{Distribution of stable self-interstitial dumbbell atoms in MoNbTaVW obtained with the tabGAP. (a) Elements making up the stable interstitial dumbbells, (b) violin plot showing distributions and average formation energies of all observed dumbbell pairs, and (c) the stable dumbbell directions.}
    \label{fig:sia_hist}
\end{figure}

To explore the preferential chemical and geometrical configurations of self-interstitial atoms, we relaxed 1000 HEA systems each containing one randomly inserted interstitial atom. Relaxation produces mainly dumbbell configurations of atom pairs. Distributions of the relaxed dumbbell configurations and formation energies are shown in Fig.~\ref{fig:sia_hist}. Fig.~\ref{fig:sia_hist}(a)-(b) reveal that there is a clear preference for V-containing interstitial dumbbells, and in particular pure V--V dumbbells which make up the majority of stable interstitials. V--V dumbbells also have on average the lowest formation energy, as shown in Fig.~\ref{fig:sia_hist}(b). Note that there are not enough non-V dumbbells to provide reliable distributions and average formation energies, but the trend of V-containing dumbbells having the lowest formation energies is clear. This can again be understood by the fact that V is the smallest atom and therefore prefers shorter interatomic bonds than the other elements.

Fig.~\ref{fig:sia_hist}(c) shows the relaxed dumbbell directions. Based on the pure elements, one would expect mainly $\hkl<111>$ directions, which are the lowest-energy configurations in Nb, V, and Ta, and directions between $\hkl<111>$ and $\hkl<221>$, which are preferred in pure Mo and W (i.e. the $\hkl<11\xi>$ interstitial revealed in Ref.~\cite{ma_symmetry-broken_2019}). Although these directions make up a large fraction of the relaxed interstitial dumbbells, Fig.~\ref{fig:sia_hist}(b) shows that \hkl<110> dumbbells are observed in almost equal numbers to \hkl<221> dumbbells. By looking at the relaxed systems, we find that the preferred direction is strongly influenced by the chemical environment. Stable configurations are found by a competition between low-energy dumbbell directions and the possibility to form V--V or other V-containing pairs. We found no statistically significant correlation between element pair and dumbbell direction, and hence only show the total distribution of directions in Fig.~\ref{fig:sia_hist}(c).

We also find that the preferential formation of V-containing dumbbell interstitials guides the mobility and migration mechanisms of single interstitials. In contrast to the pure metals, where single interstitials migrate one-dimensionally along the \hkl<111> direction in all elements~\cite{ma_symmetry-broken_2019}, interstitial migration in the HEA is three-dimensional. From constant-temperature MD simulations, we observe that the interstitials migrate mainly through connections of V atoms, so that the V-containing dumbbells is most likely to migrate to other neighbouring V atoms. If no V atoms are present in the local environment, the V interstitial remains stationary for significantly longer times.

Many of the above-discussed observations are consistent with the results by Zhao in a recent DFT study of point defects in a similar bcc alloy (VTaCrW)~\cite{zhao_defect_2020}. Namely, (1) there is no clear difference in the vacancy formation energies between the different elements, (2) the relaxation volume of vacancies directly correlates with the size of the removed atom, (3) vacancy migration energies vary significantly between the elements, (4) self-interstitial atoms are most stable when containing smaller atoms, like V--V dumbbells, and (5) self-interstitial dumbbells adopt different directions depending on the chemical environment, with $\hkl<110>$ as one of the most probable low-index directions, in clear contrast to the pure elements.

\subsection{Segregation around voids and dislocation loops}
\label{sec:cluster_segregation}

\begin{figure}
    \centering
    \includegraphics[width=\linewidth]{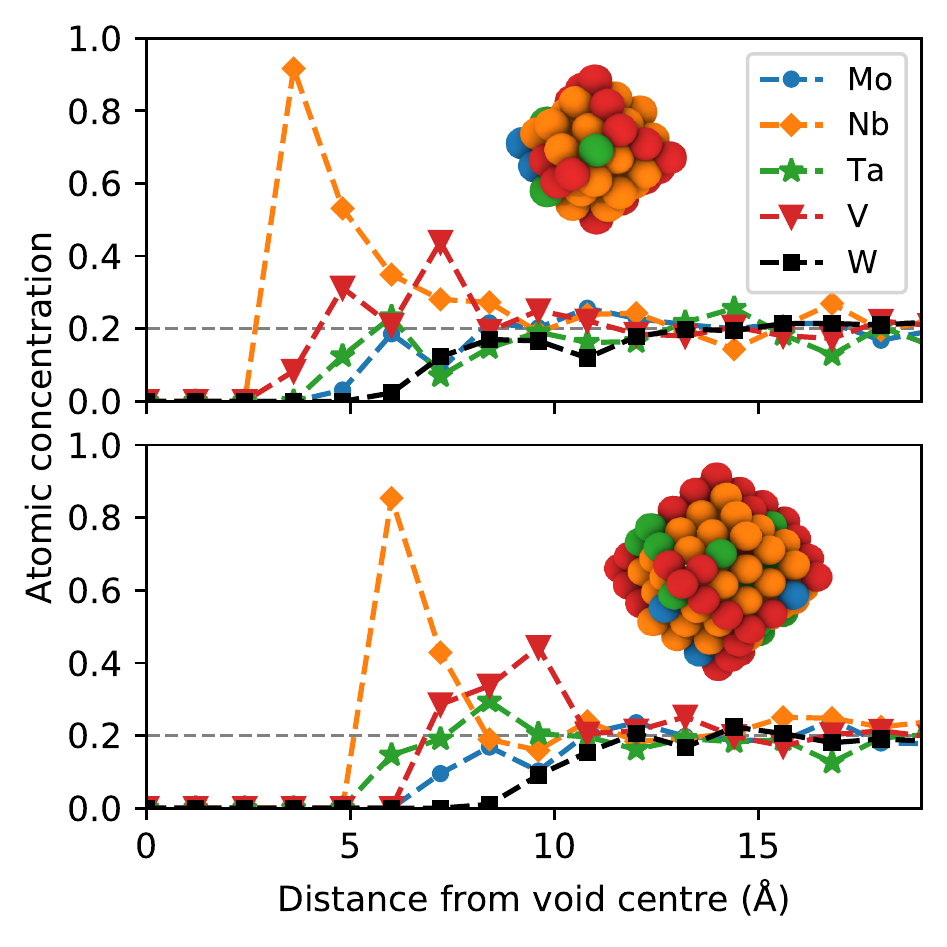}
    \caption{Segregation around voids. Concentration of each element as a function of the distance from the void centre for two different void sizes (containing 15 and 65 vacancies, corresponding to diameters of 0.8 and 1.3 nm). The snapshots show the inner surface atoms of the voids.}
    \label{fig:void}
\end{figure}

\begin{figure}
    \centering
    \includegraphics[width=\linewidth]{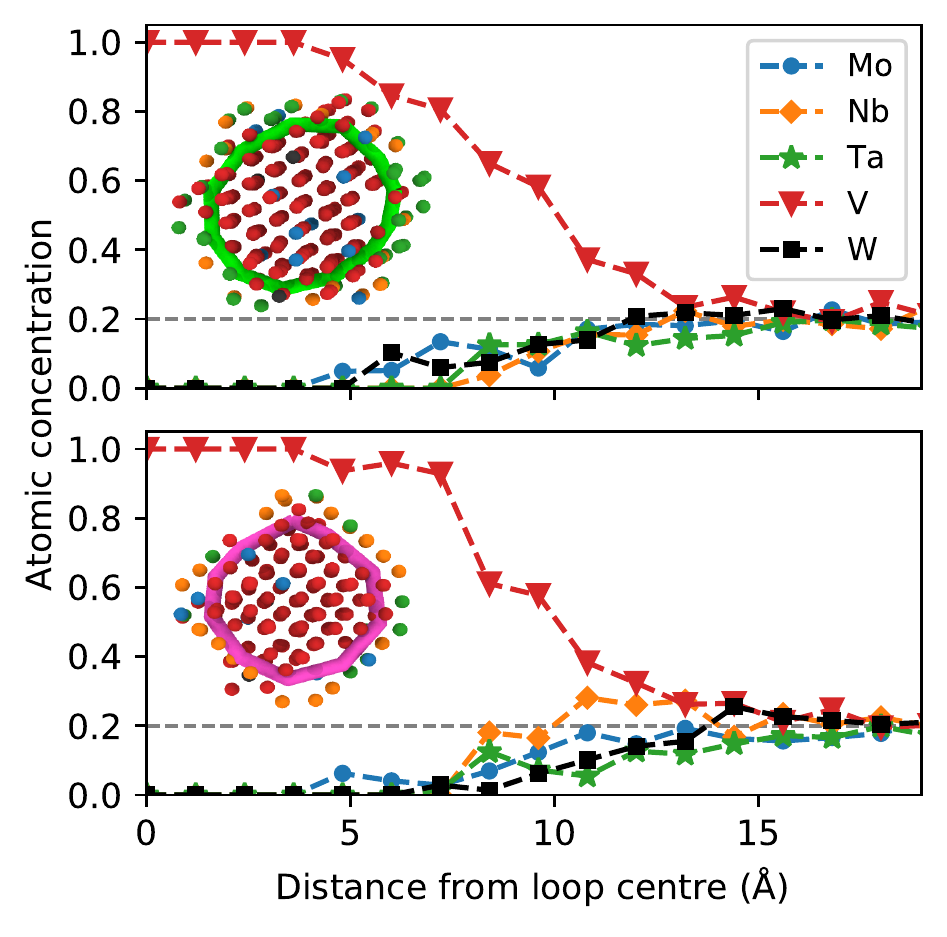}
    \caption{Segregation to interstitial dislocation loops. Concentration of each element as a function of the distance from the loop centre for a 1/2\hkl<111> loop (top) and \hkl<100> loop (bottom). The loops contains 45 and 43 interstitial atoms corresponding to a diameter of 2 nm.}
    \label{fig:loop}
\end{figure}

Going beyond single vacancies and interstitials, we now investigate how clusters of defects influence their local chemical ordering. The most stable vacancy- and interstitial-type defect clusters formed during irradiation in bcc materials are small voids and dislocation loops. We carry out MC+MD simulations for HEAs containing (separately) voids of two sizes (15 vacancies and 65 vacancies) and interstitial dislocation loops with the Burgers vectors 1/2\hkl<111> and \hkl<100> and diameter 2 nm.

The simulations show that elemental segregation around the defect clusters is clearly energetically  favoured. Fig.~\ref{fig:void} shows the concentration profiles as functions of the distance from the void centre for both void sizes. Snapshots of the inner surface atoms of the voids are shown as insets. The equilibrated inner void surface is almost completely covered by Nb. This is understandable as Nb is (along with Ta) the largest atom and also has the lowest surface energy (lower than Ta)~\cite{byggmastar_gaussian_2020}. Fig.~\ref{fig:void} also shows that there is an excess of V at the interface of the Nb-covered inner surface and the bulk. The snapshots reveal that this is because V atoms tend to decorate the edges of the void surface. The presence of the large Nb atoms and the polyhedral shape of the voids make these sub-surface regions compressed, which explains why they are preferentially occupied by V, the smallest atom.

Interstitial dislocation loops produce strong strain fields with a locally compressed lattice, which our previous observations suggest should be favoured by V. Indeed, Fig.~\ref{fig:loop} shows that essentially 100\% of the atoms within the strain field of the dislocation loops are V after the MC+MD simulations reach a steady state. This is also in line with the previous observation that the single V--V interstitial dumbbell is the most favoured interstitial configuration. The results for both voids and interstitial loops are almost identical in MC+MD simulations at both 300 K and 1000 K, showing that the preferential segregation, driven mainly by the atom size, is strong and occurs also at high temperatures.

\subsection{High-concentration defect structure}
\label{sec:annealing}

\begin{figure*}
    \centering
    \includegraphics[width=0.45\linewidth]{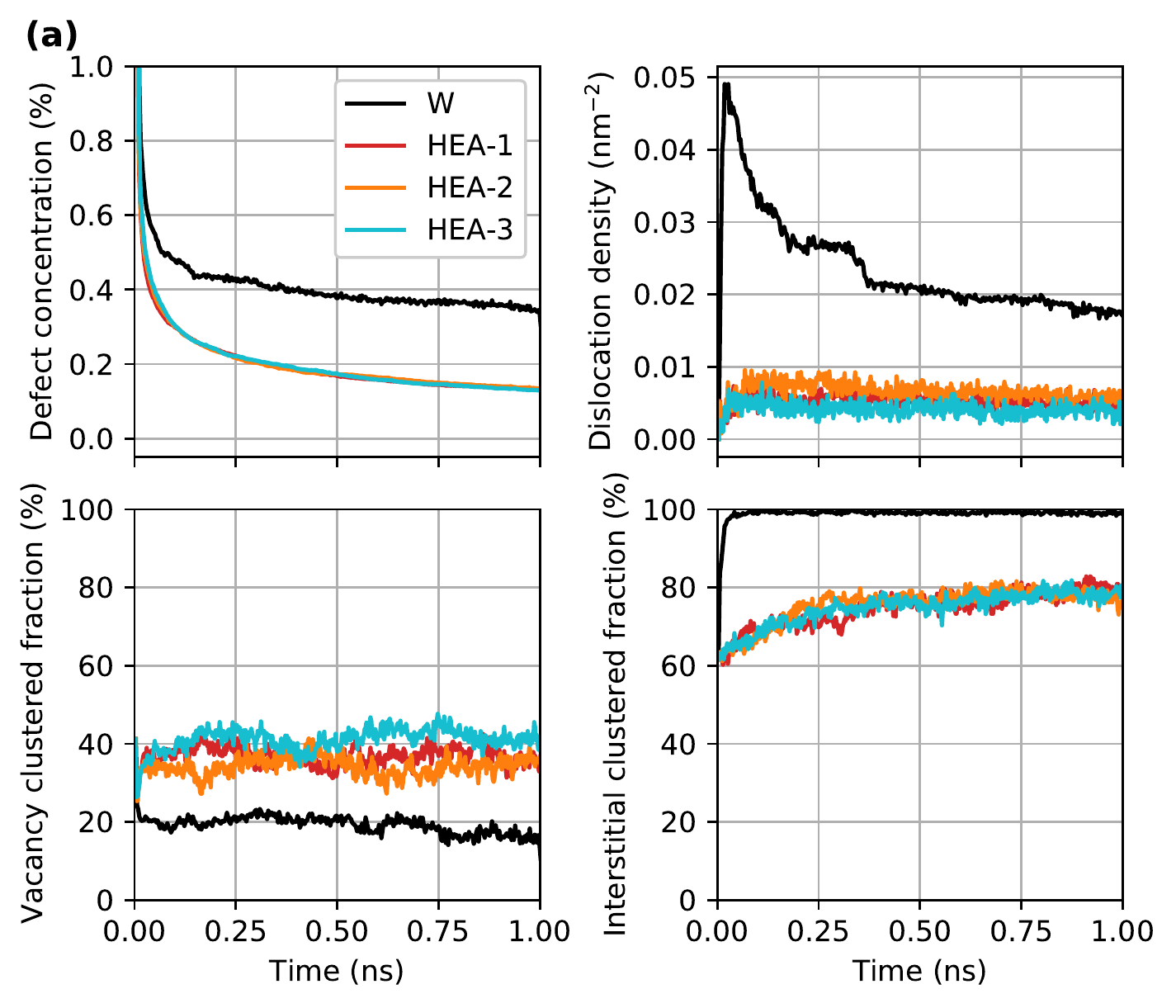}
    \includegraphics[width=0.54\linewidth]{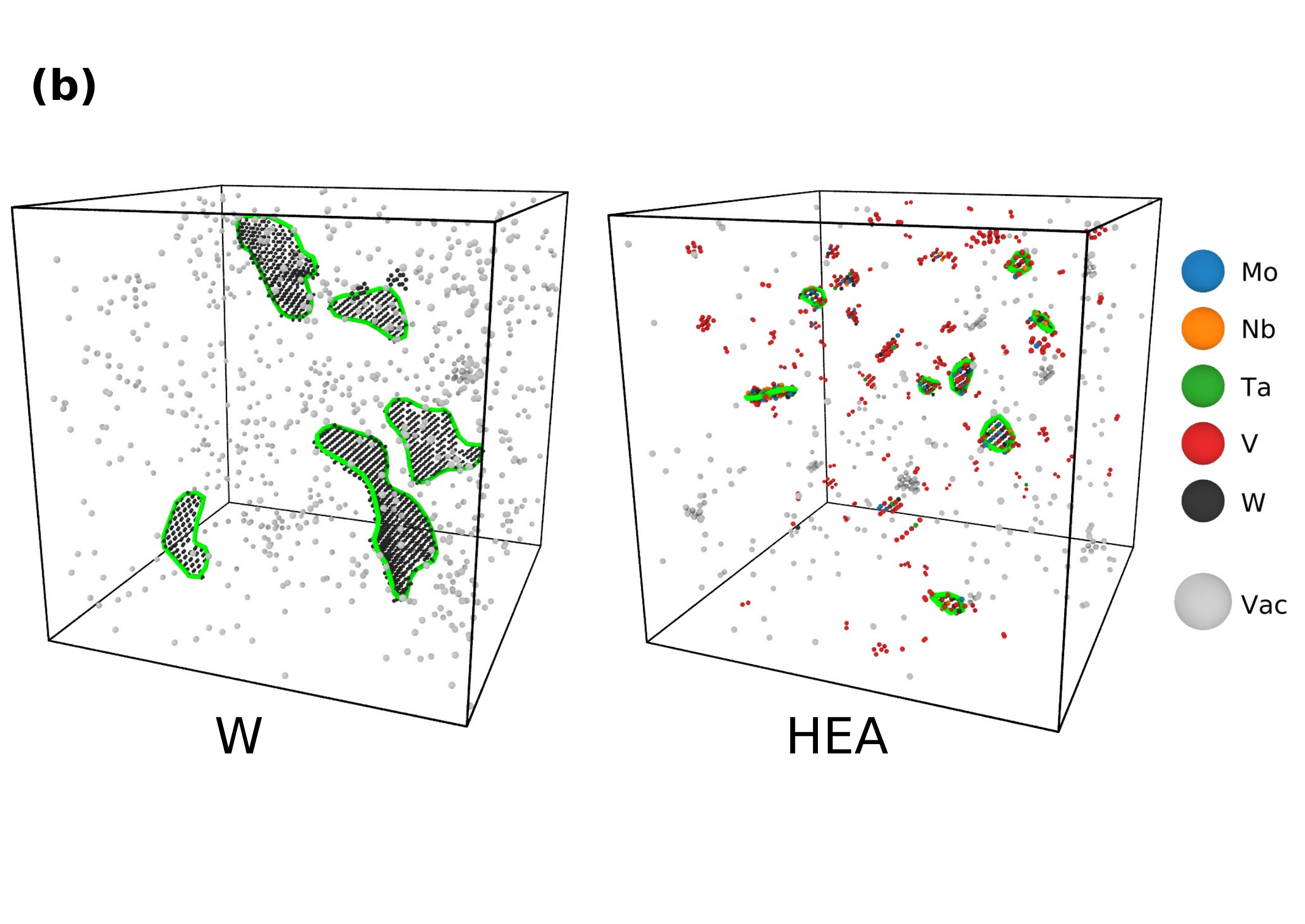}
    \caption{Defect evolution during annealing. (a) Defect statistics (defect concentration, dislocation density, and fractions of vacancies and interstitials in clusters) during annealing at 2000 K. (b) Snapshots of the final defect structure in W and the MoNbTaVW HEA, showing 1/2\hkl<111> dislocation loops (green lines), interstitial atoms, and vacancies.}
    \label{fig:anneal}
\end{figure*}

In the previous sections, we have described the energetics of point defects as well as segregation and ordering in the bulk and around possible radiation-induced defect clusters in the HEA. It is not straightforward to combine all these results into an understanding of how the HEA behaves during irradiation and the subsequent defect annealing and recovery. To this end, we carry out annealing MD simulations of a supersaturated concentration of defects and observe how the defect structure evolves over time. We prepare three 250,000-atom HEA samples with randomly inserted interstitial atoms and vacancies corresponding to a defect concentration of 4\%. After an initial minimisation of positions, the samples are annealed at zero pressure and 2000 K for 1 ns. For comparison, we also anneal a pure W sample in the same way. The results are summarised in Fig.~\ref{fig:anneal}. Animations of the defect evolution are provided in the Supplemental material online.

Fig.~\ref{fig:anneal} shows striking differences between W and the HEA during annealing. In W, interstitial atoms are extremely mobile and join to form 1/2\hkl<111> dislocation loops already during the first few picoseconds. This is clear from Fig.~\ref{fig:anneal}(a), which shows that the fraction of interstitials in clusters reaches 100\% very rapidly. Over time, the initially small dislocation loops grow by migrating and coalescing with other nearby dislocation loops, eventually resulting in only a few large loops as seen in Fig.~\ref{fig:anneal}(b). The coalescence of loops is in Fig.~\ref{fig:anneal}(a) reflected by the rapid decrease in the dislocation density. Compared to interstitials, vacancies in W migrate slowly (migration energy 1.7 eV~\cite{ma_effect_2019}). Additionally, the binding energies of small vacancy clusters in W is close to zero or even repulsive~\cite{byggmastar_machine-learning_2019}, which further limits the formation of vacancy clusters. Only around 20\% of the vacancies in W are in clusters of two or more, as shown in Fig.~\ref{fig:anneal}(a).

The defect evolution in the HEA is in many ways different. Fig.~\ref{fig:anneal}(a) shows that the recombination of defects is more efficient, especially during the early stage of the annealing simulations. Fig.~\ref{fig:anneal}(b) shows that the final defect structure in the HEA only contains small dislocation loops. These dislocation loops form already in the early part of the simulations but, unlike in W, then remain stationary and similar in size throughout the rest of the simulation. This is evidenced by the constant dislocation density over time in Fig.~\ref{fig:anneal}(a).

Contrary to W, vacancies in the HEA are more mobile (Fig.~\ref{fig:vac_mig}) and about 40\% of them are in clusters of two or more. As observed in Sec.~\ref{sec:vac_int}, single interstitials are most stable as V--V dumbbells and migrate three-dimensionally, preferentially through connections of V--V bonds. This is fundamentally different from W, where interstitials mainly migrate one-dimensionally along close-packed \hkl<111> directions with a very low migration energy. The consequence of this difference is the much more efficient defect recombination in the HEA seen in Fig.~\ref{fig:anneal}(a), as both interstitials and vacancies are mobile and can explore their surrounding through 3D migration. Furthermore, because the dislocation loops are unable to move and interstitials are more likely to recombine with nearby vacancies, the overall defect structure does not significantly change over time. In particular, the dislocation loops remain very small. All loops are around 1--1.5 nm in diameter with the Burgers vector 1/2\hkl<111>, with most of the loops around 1 nm and consisting of 20--30 interstitials. Smaller interstitial clusters are also mostly parallel \hkl<111> interstitials, but cannot be classified as dislocation loops and are thus not identified by the dislocation extraction algorithm. Additionally and in line with our observations in Secs.~\ref{sec:vac_int} and \ref{sec:cluster_segregation}, all interstitial clusters are enriched by V.

\section{Discussion}

We have studied segregation and the stability and evolution of radiation-induced defects in MoNbTaVW using a new machine-learned interatomic potential. The potential relies on low-dimensional two- and three-body descriptors, which allows for good accuracy with a moderately sized training data set. It also allows for creating tabulated potentials, which can be efficiently evaluated using cubic spline interpolations without carrying out the underlying machine-learning regression. The resulting machine-learned spline potential (the tabGAP) thus retains the flexibility of the machine learning framework, but runs at a speed comparable to conventional analytical three-body potentials. Given the accuracy achieved, our work demonstrates that developing low-dimensional tabulated machine-learned potentials is a promising alternative or complement to more expensive many-body machine-learning potentials~\cite{bartok_gaussian_2010,shapeev_moment_2016,behler_generalized_2007,thompson_spectral_2015}, especially for multicomponent alloys where data efficiency becomes increasingly important.

Through MD simulations with the machine-learned potential, we have shown that there is a clear preferential chemical ordering in the bulk MoNbTaVW HEA, with mainly Mo--Ta, V--W, and Mo--Nb binaries, in agreement with previous studies~\cite{fernandez-caballero_short-range_2017-1,kostiuchenko_impact_2019,liu_monte_2021}. Our simulations also revealed that the presence of defects introduces strong traps for elemental segregation. We observe strong segregation of Nb to spacious regions like voids. The inner surfaces of voids are preferentially almost completely covered by Nb, due to Nb being the largest atom with the lowest surface energy. We also found that V as the smallest atom prefers compressed regions such as the strain field of interstitial dislocation loops or even single interstitial dumbbell configurations.

It is worth noting that for the observed segregation and ordering of elements to take place, there must occur a mass transport of the given elements through favourable migration mechanisms. In our MC+MD simulations, the kinetics of the segregation and ordering is not explicitly modelled. However, the observations in Section~\ref{sec:vac_int} bring additional insight. Our results showed that the vacancy migration energies are lowest for Nb, Ta, and V, which offers a possible pathway for the observed segregation of Nb and V to defect clusters. Additionally, we found that single interstitials occur mostly as V-containing dumbbells that migrate preferentially by connecting with a neighbouring V atom, resulting in mass transport of V through interstitial diffusion. In the annealing simulations in Section~\ref{sec:annealing} we observe this directly as the final dislocation loops and small interstitial clusters are all enriched by V. 

Perhaps the most striking result is the observation in Sec.~\ref{sec:annealing} that dislocation loops never grow to substantial sizes when annealing MoNbTaVW containing a large concentration of defects. This results in a radically different response to irradiation than pure bcc metals like W. Although it still remains unclear how energetic radiation-induced collision cascades would affect the defect evolution (i.e. whether larger clusters could be directly produced in cascades), our results offer an explanation for recent experimental observations. El-Atwani et al.~\cite{el-atwani_outstanding_2019} recently irradiated a W-based Cr--Ta--V--W alloy up to a high dose of 8 dpa and yet did not resolve any dislocation loops from the transmission electron microscopy (TEM) images, observing only black spots. In comparison, TEM-visible dislocation loops form in pure W already at extremely low dose ($\leq 0.01$ dpa) as a result of individual high-energy collision cascades~\cite{yi_-situ_2016-1,sand_high-energy_2013}. Using atom probe tomography and cluster-expansion modeling, it was concluded that the black spots were precipitates of Cr and V~\cite{el-atwani_outstanding_2019}. Our results suggest that even without such a second-phase precipitation (which we do not observe in MoNbTaVW) radiation-induced dislocation loops still remain very small (around 1 nm). Considering the strain that comes from different atom sizes of the constituent elements, such very small dislocations are not fully ordered, and hence likely to be invisible or seen only as black spots in TEM images.

\section{Summary}

In summary, we have developed a computationally fast machine-learned potential for Mo--Nb--Ta--V--W alloys and used it to advance the understanding of segregation and defect structure in the equiatomic MoNbTaVW high-entropy alloy. We observed clear segregation of Nb to vacancy-rich spacious regions and V to interstitial-rich compressed regions. Furthermore, our results indicate that dislocation loops remain very small in irradiated MoNbTaVW and most likely experimentally invisible or seen as structureless 'black spot' damage. We explain this by the comparable and three-dimensional migration of interstitials and vacancies combined with the immobility of the small dislocation loops, which leads to more efficient defect recombination at the expense of clustering. Our work opens up possibilities for further studies on the radiation-induced chemical ordering and defect structure and paves the way for a computational search of promising high-entropy  Mo--Nb--Ta--V--W compositions beyond the equiatomic alloy.





\section*{Acknowledgements}

This work has been carried out within the framework of the EUROfusion consortium and has received funding from the Euratom research and training programme 2014-2018 and 2019-2020 under grant agreement No 633053. The views and opinions expressed herein do not necessarily reflect those of the European Commission. Grants of computer capacity from CSC - IT Center for Science, Finland, as well as from the Finnish Grid and Cloud Infrastructure (persistent identifier urn:nbn:fi:research-infras-2016072533) are gratefully acknowledged.





\bibliography{mybib}

\end{document}


\title{Supplemental material for: Modeling refractory high-entropy alloys with efficient machine-learned interatomic potentials: defects and segregation}

\author{J. Byggmästar}
\thanks{Corresponding author}
\email{jesper.byggmastar@helsinki.fi}
\affiliation{Department of Physics, P.O. Box 43, FI-00014 University of Helsinki, Finland}
\author{K. Nordlund}
\affiliation{Department of Physics, P.O. Box 43, FI-00014 University of Helsinki, Finland}
\author{F. Djurabekova}
\affiliation{Department of Physics, P.O. Box 43, FI-00014 University of Helsinki, Finland}
\affiliation{Helsinki Institute of Physics, Helsinki, Finland}

\date{\today}

\maketitle

\section{Training and testing data}

The Mo--Nb--Ta--V--W training data set is available from Ref.~\cite{byggmastar_ida_tabgap_2021} and consists of the following structures:

\begin{itemize}
    \item Single isolated atoms to get the correct cohesive energy.
    \item Dimers, mainly to guide the repulsive potential fit as described in Ref~\cite{byggmastar_machine-learning_2019}.
    \item Pure elements. The pure-element structures are small subsets of the training data used in Ref~\cite{byggmastar_gaussian_2020}. To sample representative subsets of the different pure-element structure types (bulk, vacancy, liquids, etc.) we used farthest-point sampling of the average SOAP vectors~\cite{bartok_representing_2013}, as described in Ref.~\cite{de_comparing_2016}.
    \item Binary bcc alloys. For each element pair $A$--$B$ we sampled 10 different concentrations from $A_{0.05}B_{0.95}$ to $A_{0.95}B_{0.05}$ and 3 different lattice constants for every composition. Atoms are randomly ordered and shifted slightly from their lattice positions.
    \item Ternary, quaternary, and quinary bcc alloys. For every alloy combination we sampled 3 linearly spaced compositions and 3 different lattice constants. Atoms are randomly ordered and shifted slightly from their lattice positions.
    \item Bulk equiatomic quinary HEAs. Atoms are randomly ordered and shifted slightly from their lattice positions. The lattice constant is randomised in the range 3--3.4 Å.
    \item Various ordered binary, ternary, and quaternary alloys (always as a bcc lattice, but with different crystal symmetries of the elemental sublattices). Some equiatomic quinary alloys were subsequently added in an active-learning fashion by running MC+MD simulations with the previously trained tabGAP version. The energies and forces of the resulting partially ordered alloys were then computed with DFT and added to the training set. After retraining the potential with the new and extended training data set, the process was repeated until the accuracy for the next batch of MC+MD-predicted alloys was similar to the other test errors (a few meV/atom). Only two iterations were needed to reach this convergence.
    \item Liquid equiatomic binary, ternary, quaternary, and quinary alloys at different densities, with most data for the quinary equiatomic HEA. The structures were prepared by running MD with different generations of the tabGAP, initially trained to only crystal structures, and then iteratively trained to more and more liquids similar to the description above.
    \item Vacancies in HEAs. 1--5 vacancies were randomly inserted in HEA lattices and then optimised or relaxed in constant-temperature MD simulations with the (at the time) current version of the tabGAP. Structures were iteratively added in the active-learning approach described above. 
    \item Interstitial atoms in HEAs. 1--5 interstitial atoms were randomly inserted in HEA lattices and then optimised with the (at the time) current tabGAP version. Structures were iteratively added in the active-learning approach described above. 
    \item Disordered HEA surfaces. Some of the damaged/molten surface structures from the pure W training data~\cite{byggmastar_machine-learning_2019} turned into HEAs.
    \item Short-range interstitial atoms. Randomly placed unrelaxed interstitial atom in HEAs to fit repulsion inside crystals, making sure that the closest interatomic distance is not too short for DFT to be unreliable ($> 1.35$ Å).
\end{itemize}

Most of the bcc structures described above contain 54 atoms ($3\times3\times3$ unit cells). Liquid, vacancy, interstitial, and the MC+MD-ordered structures contain 128 atoms. Smaller test sets were prepared for the most relevant classes of structures.

\section{Additional test-set errors}

\begin{figure}[h]
    \centering
    \includegraphics[width=\linewidth]{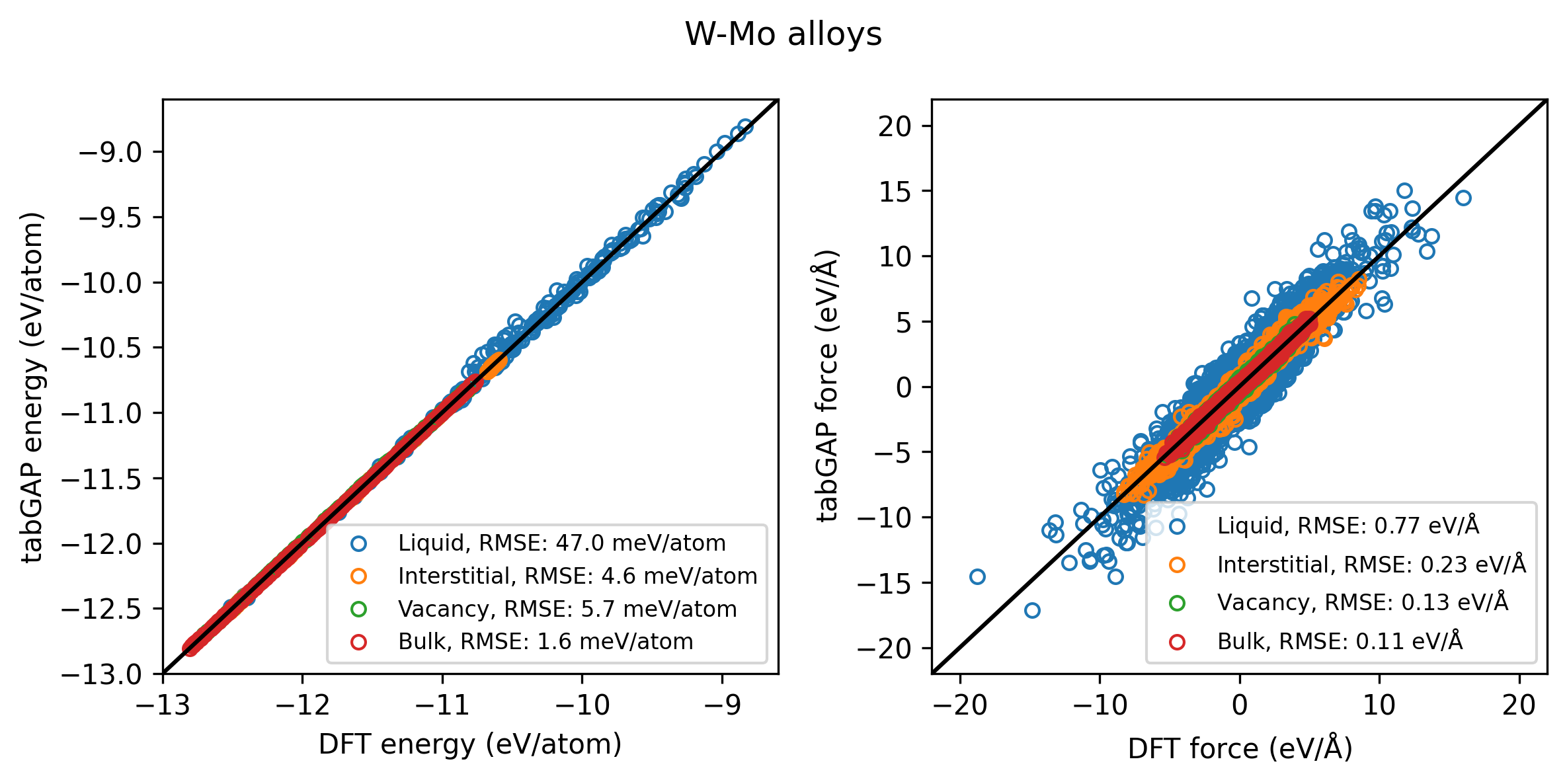}
    \caption{Accuracy of the tabGAP for binary W--Mo alloys, using parts of the training data from Ref.~\cite{nikoulis_machine-learning_2021} as test sets.}
    \label{fig:W-Mo}
\end{figure}

Fig.~\ref{fig:W-Mo} shows the comparison between the tabGAP and sets of W--Mo structures from the DFT training data of Ref.~\cite{nikoulis_machine-learning_2021}. The data set contains W--Mo alloys across the entire composition range as liquids, bcc lattices with single interstitial atoms or single vacancies, and bulk bcc lattices. The root-mean-square errors (RMSE) are similar to the test errors shown in the main article, indicating that the potential is accurate also for binary alloys.

\begin{figure}[h]
    \centering
    \includegraphics[width=0.45\linewidth]{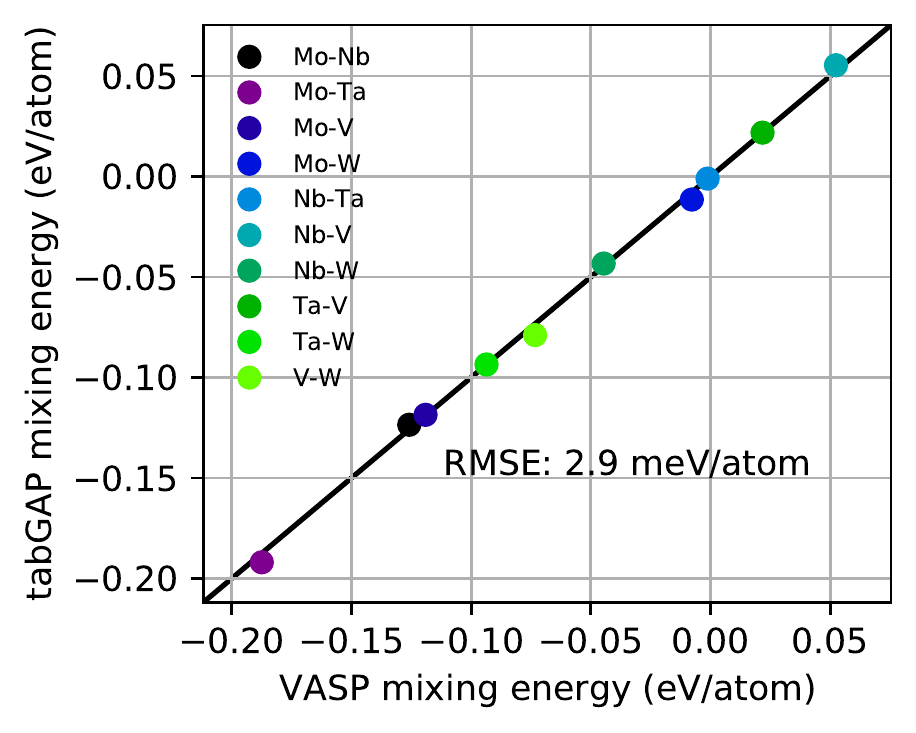}
    \caption{Mixing energies of CsCl-ordered (B2) binary alloys compared between DFT and the tabGAP.}
    \label{fig:B2}
\end{figure}

Fig.~\ref{fig:B2} shows the mixing energies of all binary alloys ordered in the CsCl (B2) symmetry. The lattices are fully relaxed in both VASP and with the tabGAP. The agreement between the two is good, with an RMSE of 2.9 meV/atom for the mixing energy.

\clearpage
\section{Three-body grid size convergence}

\begin{figure}[h]
    \centering
    \includegraphics[width=0.8\linewidth]{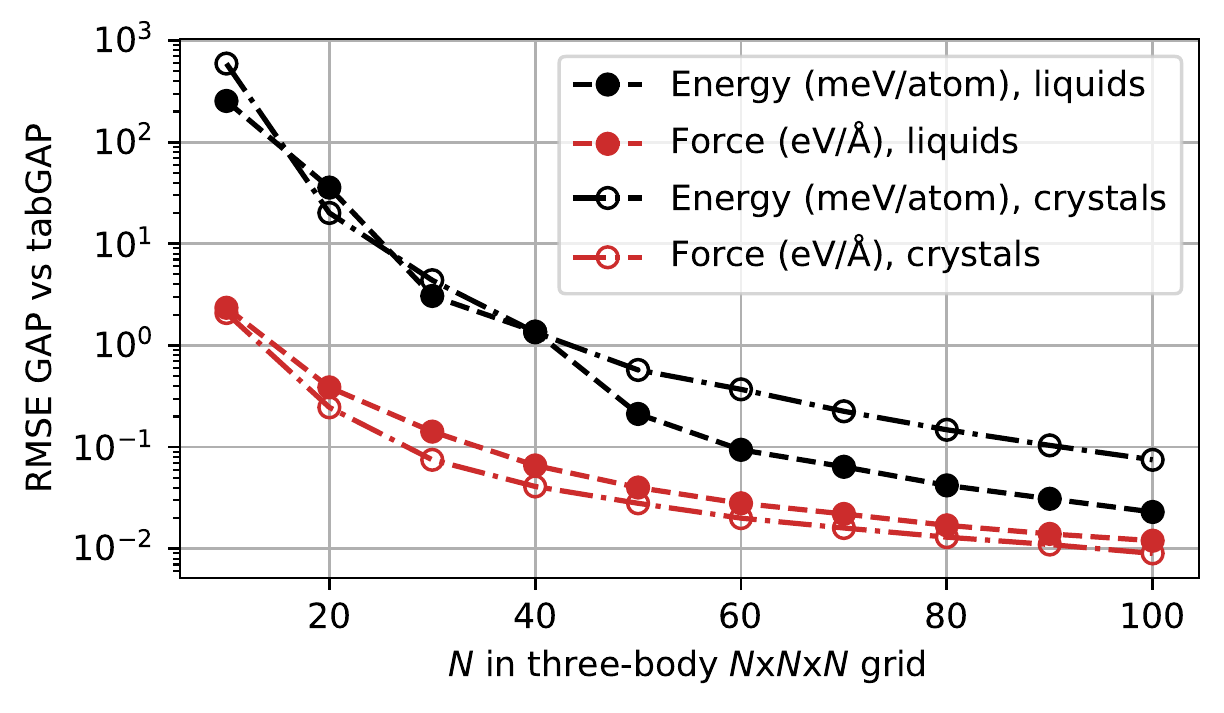}
    \caption{Convergence of the three-body ($r_{ij}$, $r_{ik}$, $\cos \theta_{ijk}$) grid tested for two data sets of crystalline and liquid structures containing both alloys and pure elements. The tabGAP used in all simulations is tabulated using $N=80$.}
    \label{fig:grid_convergence}
\end{figure}

Fig.~\ref{fig:grid_convergence} shows the convergence of the error of the tabGAP compared to the original GAP as a function of the size of the 3D grid for the three-body interpolation. Convergence curves are shown for structures from the training data, containing liquid and crystalline structures and both pure elements and all alloy compositions. The ($r_{ij}$, $r_{ik}$, $\cos \theta_{ijk}$) grid is sampled from $r_{ij} = r_{ik} = 0.1$ Å to 5 Å (the cutoff distance) and $\cos \theta_{ijk}$ from $-1$ to 1. For the final tabGAP we used an $80\times80\times80$ grid, for which the interpolation errors are negligible ($\sim 0.1$ meV/atom, $\sim 0.01$ eV/Å) and around ten times lower than the accuracy of the tabGAP/GAP when compared to DFT ($\geq1$ meV/atom, $\geq 0.1$ eV/Å). For the tabGAP pair potential parts we used 5000 points between 0.02 and 5 Å, for which the interpolation error is vanishingly small.

\clearpage
\section{Machine-learned pair potentials}

\begin{figure}[h]
    \centering
    \includegraphics[width=\linewidth]{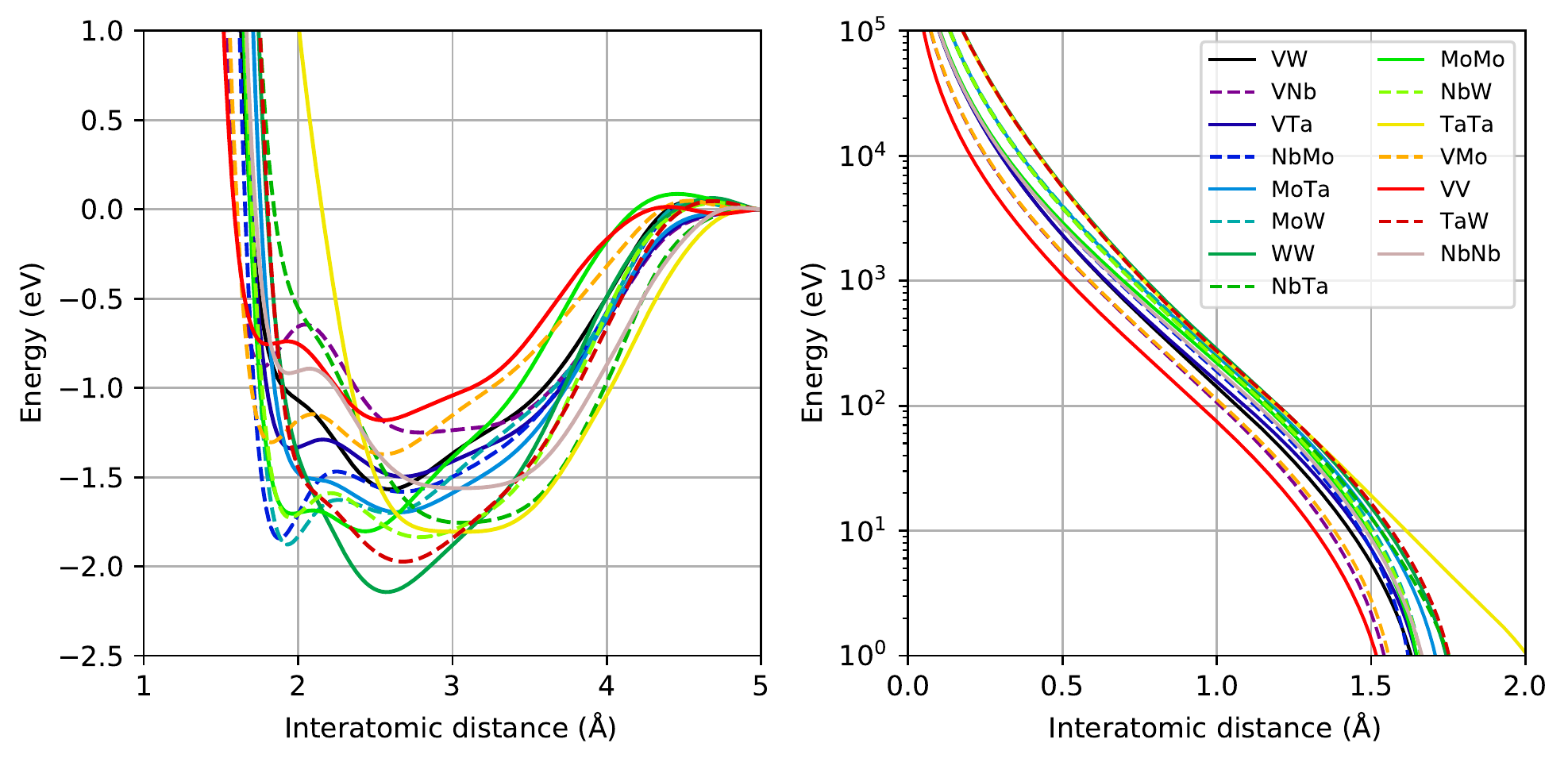}
    \caption{The machine-learned pair potentials, showing the near-equilibrium parts (left) and the short-range repulsive parts (right).}
    \label{fig:pairpots}
\end{figure}

Fig.~\ref{fig:pairpots} shows the machine-learned pair potentials for every element combination, verifying that all potentials are smooth and well-behaved. Note that we did not require the near-equilibrium parts of the potentials to follow the realistic dimer dissociation curves, even though the training data does contains all dimers. Dimers were mainly included to guide the repulsive parts and the near-equilibrium pair potentials were allowed to be adjusted to best reproduce the bulk structures. This is evident in Fig.~\ref{fig:pairpots} by the fact that most pair potentials show two minima, one close to the nearest-neighbour distance of the bulk and the other close to the dimer bond length. However, we stress that this does not result in any spurious behaviour in the bulk where many bonds and the three-body potentials make up the total energy.

\clearpage
\section{Machine-learned three-body potentials}

Fig.~\ref{fig:3b} shows a few examples of the machine-learned three-body potentials. The energy landscape is illustrated as the energy contribution to atom $i$ in a triplet $ijk$, where $r_{ij}$ is fixed to the nearest-neighbour distance in the bcc crystal ($\sim 2.77$ Å). Fig.~\ref{fig:3b} verifies that the three-body potentials produce a smooth energy landscape. They also show that the majority of the energy is contributed by the first- and second-nearest neighbour atoms as the regions where $r_{ik} \gtrsim 3.2$ Å and $\theta_{ijk} \gtrsim 100$ degrees are relatively flat. Since the shapes of the three-body potentials are fully machine-learned without restrictions, this can be taken as a justification for the fact that analytical many-body potentials for bcc metals are often restricted to the second-nearest-neighbour interactions. Nevertheless, we found that a 5 Å three-body cutoff (which includes the third-nearest neighbours) still provides a significant enough improvement in the overall accuracy.

\newcommand\myW{0.32}

\begin{figure}[h]
    \centering
    \includegraphics[width=\myW\linewidth]{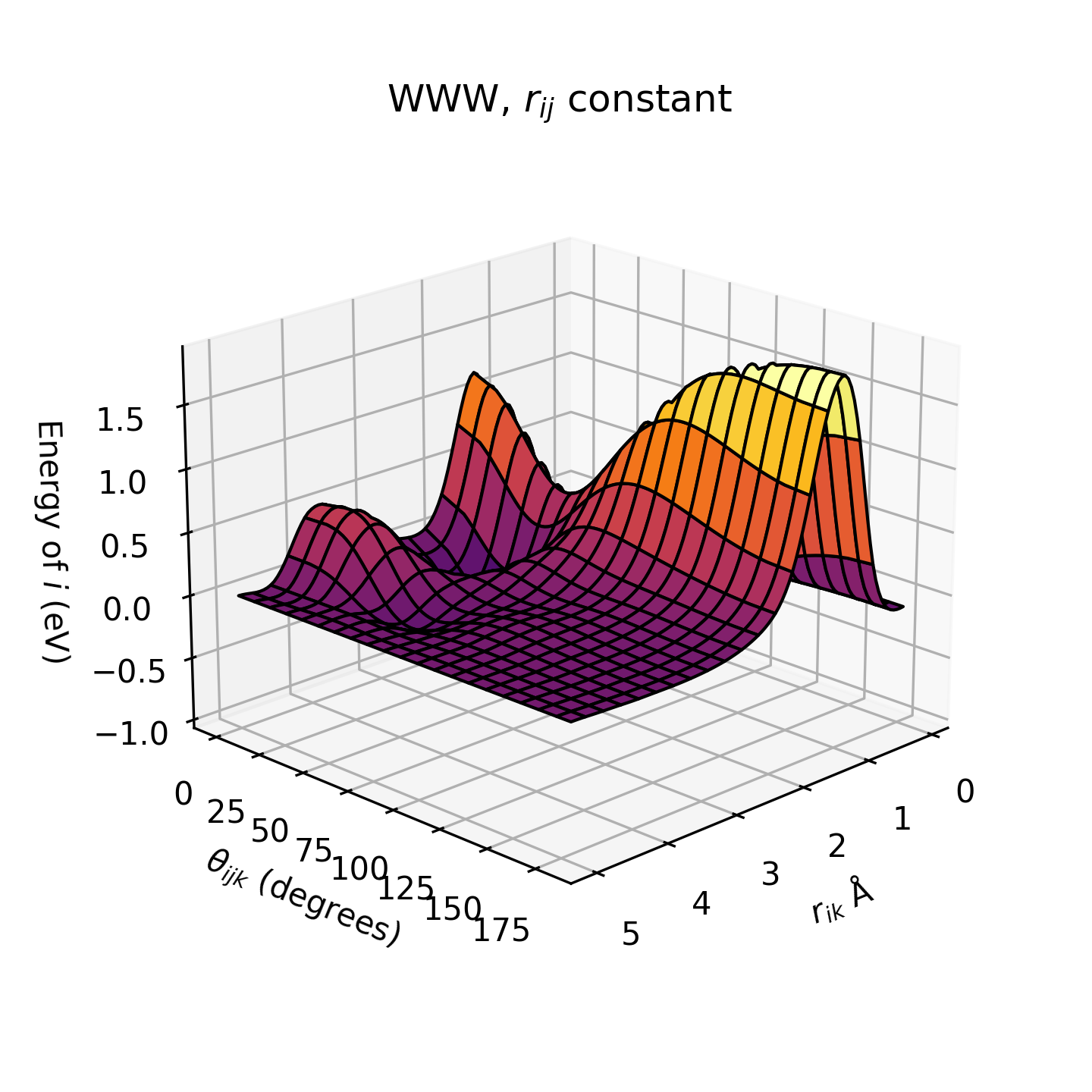}
    \includegraphics[width=\myW\linewidth]{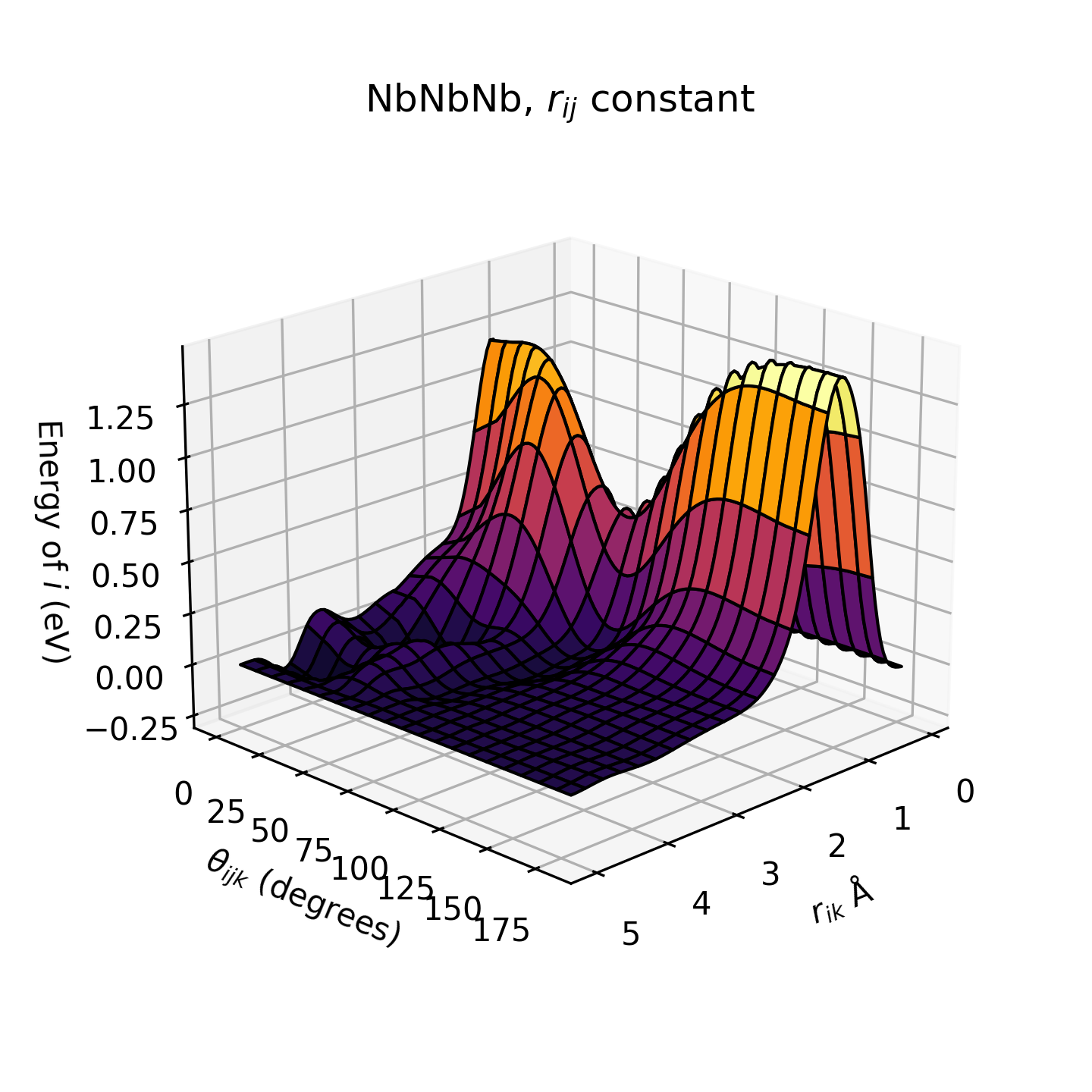}
    \includegraphics[width=\myW\linewidth]{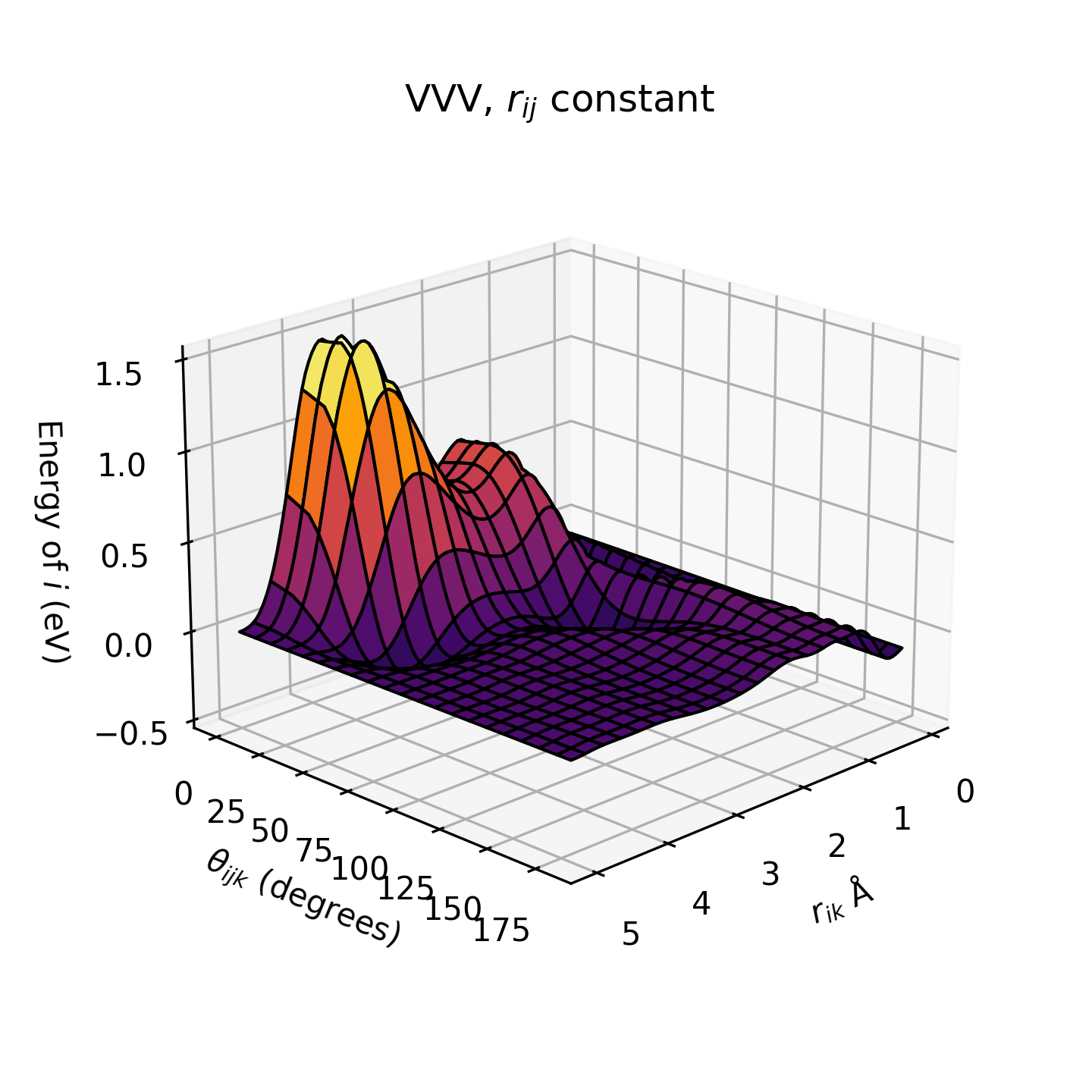}
    \includegraphics[width=\myW\linewidth]{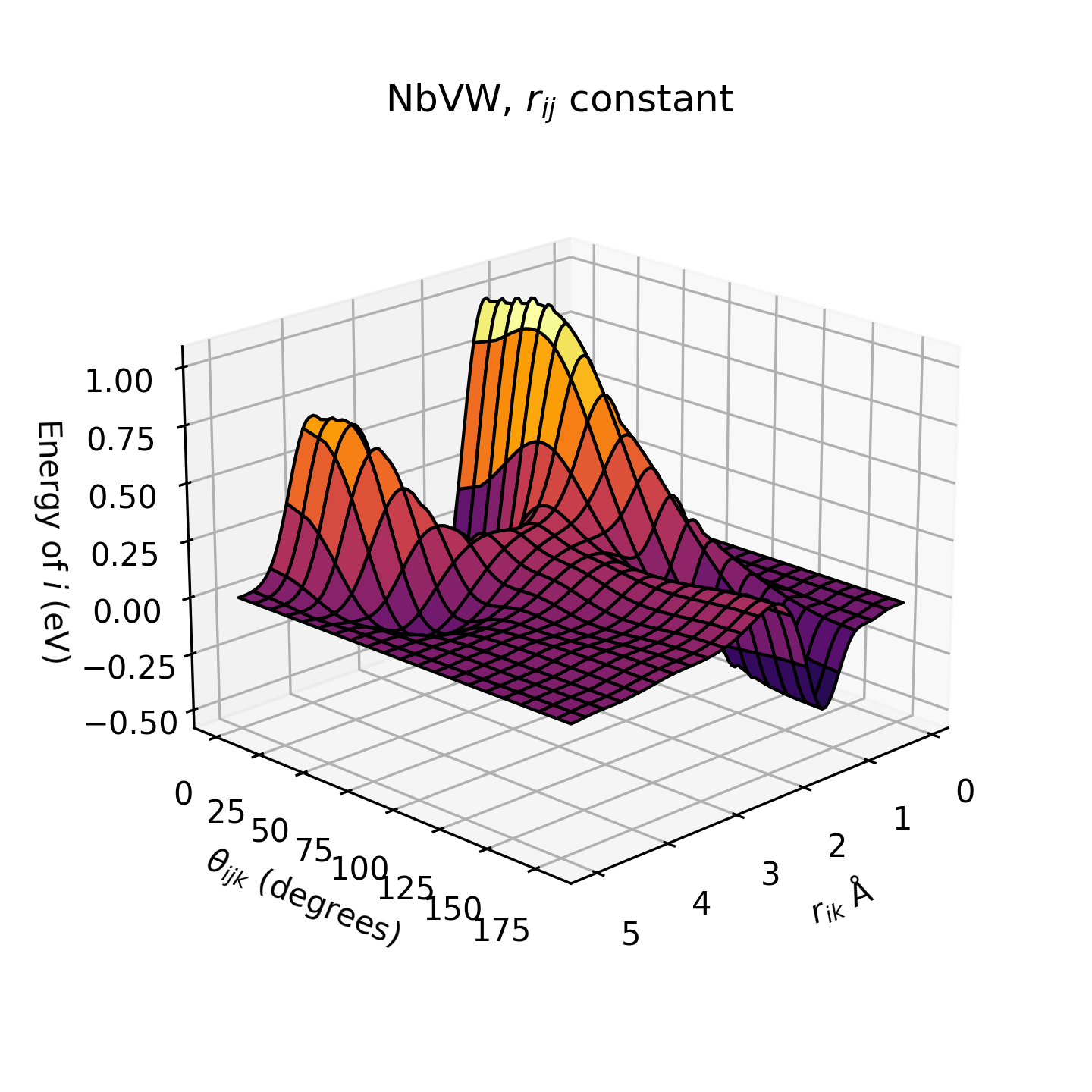}
    \includegraphics[width=\myW\linewidth]{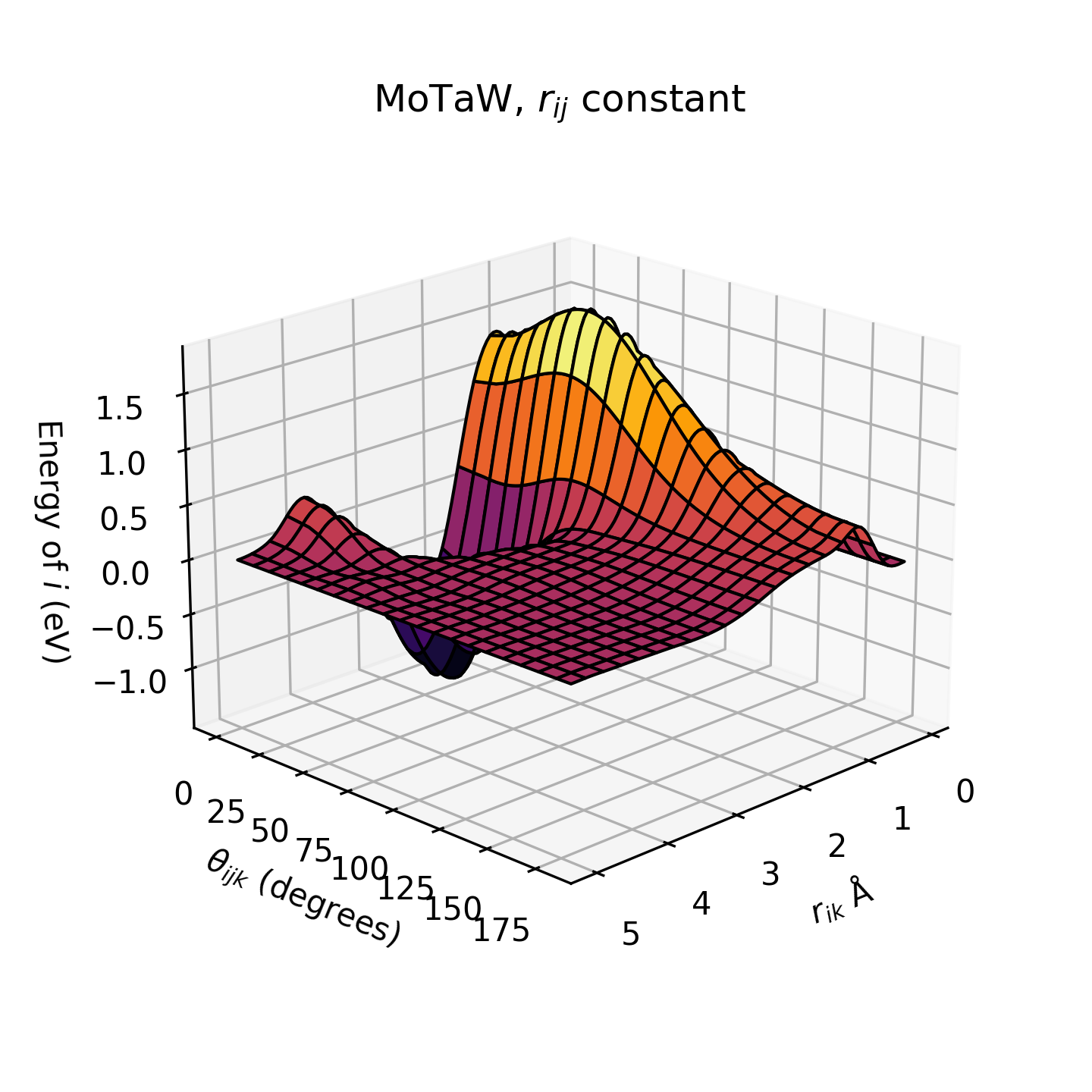}
    \includegraphics[width=\myW\linewidth]{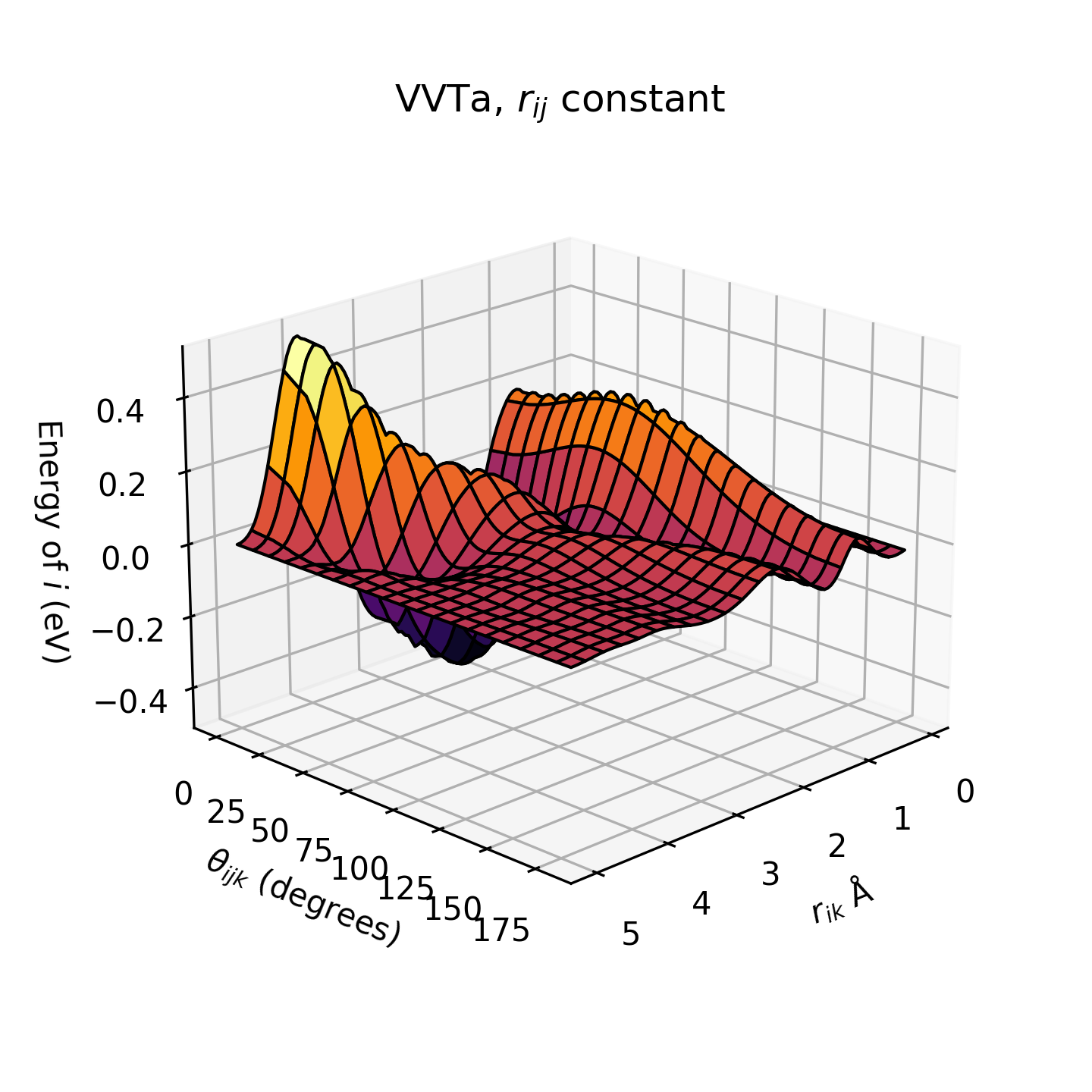}
    \caption{Examples of the machine-learned three-body potentials, visualised with $r_{ij}$ fixed to the nearest-neighbour distance in bcc ($\sim 2.77$ Å). The vertical axis is the three-body energy of atom $i$ in the triplet $ijk$.}
    \label{fig:3b}
\end{figure}

\bibliography{mybib}